\documentclass[12pt,notitlepage,showkeys,noeprint,superscriptaddress]{revtex4-1}
\usepackage{bbm}
\usepackage[colorlinks=true,
colorlinks=true
]{hyperref}
\usepackage{graphicx}
\usepackage{subfigure}
\usepackage[T1]{fontenc}

\usepackage{multirow}

\begin{document}

\title{Bell-Mermin-Klyshko Inequalities and One-way Information Deficit of Dirac Fields in Noninertial Frames}
\author{Biao-Liang Ye}
\email{biaoliangye@gmail.com}
\affiliation{Quantum Information Research Center, Shangrao Normal
University, Shangrao 334001, China}

\author{Yao-Kun Wang}
\affiliation{College of Mathematics, Tonghua Normal University, Tonghua, Jilin 134001, China}

\author{Shao-Ming Fei}
\email{feishm@cnu.edu.cn}
\affiliation{School of Mathematical Sciences, Capital Normal University, Beijing 100048, China}
\affiliation{Max-Planck-Institute for Mathematics in the Sciences, 04103 Leipzig, Germany}

\begin{abstract}
We investigate the Bell-Mermin-Klyshko inequalities and the one-way
information deficit of Dirac fields in noninertial frames, where the quantum correlations
are shared between inertial and accelerated observers due to the Unruh effect. We derive partial analytical results for specific quantum states using the one-way information deficit. Additionally, we present numerical results for the Bell-Mermin-Klyshko inequalities. The study reveals the presence of Bell nonlocality and the significance of the one-way information deficit in relativistic quantum information.
\end{abstract}

\maketitle
\date{\today}

\section{Introduction}
Quantum information science has received
significant attention due to its crucial
role in various applications such as
quantum teleportation \cite{Bennett1993},
remote state preparation \cite{Bennett2001},
quantum key distribution \cite{Scarani2009} and
quantum secure direct communication \cite{Wang2005}. In particular,
the quantum entanglement in the relativistic frame has
attracted considerable interest \cite{Peres2004}. It has been shown that
entanglement plays a crucial role in understanding
black hole \cite{Fuentes-Schuller2005,Martin-Martinez2009,Martin-Martinez2010b} and noninertial frame \cite{Alsing2006,Martin-Martinez2011,
Hwang2011,Friis2011,Montero2011,Chang2012,
Montero2012,Wang2020} to illustrate diverse properties in space time \cite{Alsing2003,Martin-Martinez2010,Martin-Martinez2010a,Bruschi2010,Montero2011a,Ahmadi2016,Xu2020,Li2022,Liu2023,Zhang2023}.
The decoherence \cite{Wang2010a} and recover \cite{Xiao2018} of entanglement have been also extensively studied.

Beyond entanglement, other quantum correlations
such as quantum discord \cite{Ollivier2001},
measurement-induced disturbance \cite{Luo2008}, geometric discord \cite{Dakic2010},
one-way information deficit \cite{Horodecki2005,Streltsov2011,Ciliberti2013,Wang2013,Wang2015,Ye2017} play an essential role in quantum information
processing \cite{Modi2012,Adesso2016}. The investigation of quantum correlations beyond
quantum entanglement in relativistic quantum information has
become a hot topic \cite{Datta2009,Wang2010,Mehri-Dehnavi2011,Wang2014}. Quantum discord
has been employed to explore the Unruh Hawking effect, indicating that
the initial entanglement between two subsystems degrades when one subsystem undergoes uniform acceleration \cite{Fuentes-Schuller2005}. In a recent study \cite{Wu2022}, two quantum
correlation approaches were used to analyze the physically accessible and inaccessible observers caused by Hawking radiation under the Schwarzschild spacetime. Haddadi et al. \cite{Haddadi2024} utilized quantum correlations, local quantum uncertainty
and local quantum Fisher information to examine the quantumness near a
Schwarzschild black hole under decoherence. Nevertheless,
the one-way information deficit has not been studied in relativistic
quantum information theory, particularly in
noninertial quantum systems. In this
paper, we also consider the Bell inequalities \cite{Bell1964} and their generalization to Dirac fields
in noninertial frames. The Mermin inequalities which represent a form of generalized Bell inequalities, have already been investigated in quantum field theory by the group of De Fabritiis et al. \cite{DeFabritiis2023}. Furthermore, additional related works in this direction can be found in \cite{DeFabritiis2024,Guimaraes2024}. 

The paper is structured as follows. In Section II, we provide
an overview of the fundamental concepts concerning Dirac
particles from a uniform accelerating observer, the Bell-Mermin-Klyshko inequalities and one-way information deficit. The main results are presented in Section III. We summarize in Section IV.

\section{The model, Bell-Mermin-Klyshko inequalities and one-way information deficit}
{\bf The model of Dirac fields} The Minkowski Dirac field
$\psi$ satisfies \cite{Datta2009,Wang2010}
\begin{eqnarray}
i\gamma^\mu\partial_\mu\psi-m\psi=0,
\end{eqnarray}
where $\gamma_\mu$ is the Dirac gamma operator, $m$ denotes the mass of particle.
The field may be derived from the inertial and uniformly accelerated
observers \cite{Alsing2006}.

The model we investigate including an inertial observer
named Alice and a uniformly accelerating
observer Rob \cite{Wang2010a,Xiao2018}.
For the inertial observer, under Minkowski
coordinates the quantized field can be enlarged in
terms of a complete set of modes
\begin{eqnarray}
\psi=\int dk(a_\mathbf{k}\psi_\mathbf{k}^++b_\mathbf{k}^\dagger\psi_\mathbf{k}^-),
\end{eqnarray}
where $\psi_\mathbf{k}^+$ and $\psi_\mathbf{k}^-$ are
the positive and negative frequency modes, $\mathbf{k}$ is the wave vector to label
the modes of massive Dirac fields,
$a_\mathbf{k}$ and $b_\mathbf{k}$ ($a_\mathbf{k}^\dagger$ and $b_\mathbf{k}^\dagger$) are the annihilation (creation) operators for the frequency modes of momentum $\mathbf{k}$ satisfying
$\{a_\mathbf{k},a_{\mathbf{k}'}^\dagger\}=\{b_\mathbf{k},b_{\mathbf{k}'}^\dagger\}
=\delta(\mathbf{k}-\mathbf{k}')$.

The uniform accelerating observer Rob's coordinates are described by the
noninertial Rindler coordinates \cite{Alsing2003,Wang2010a}. Rob travels on
a hyperbola limited in the Rob region I which will trigger disconnecting
from region II. The transformation from the Minkowski coordinates $(t,z)$
to Rob coordinates $(\tau, \xi)$ is given by
$t=e^{a\xi}\sinh(a\tau)/a$ and $z=e^{a\xi}\cosh(a\tau)/z$ with $a$ being the Rob's acceleration. And the Dirac field in terms of positive
and negative frequency modes is denoted as
\begin{eqnarray}
\psi=\int dk(c_\mathbf{k}^I\psi_\mathbf{k}^{I+}+d_\mathbf{k}^{I\dagger}\psi_\mathbf{k}^{I-}
+c_\mathbf{k}^{II}\psi_\mathbf{k}^{II+}+d_\mathbf{k}^{II\dagger}\psi_\mathbf{k}^{II-}),
\end{eqnarray}
where $(c_\mathbf{k}^r,c_\mathbf{k}^{r\dagger})$ and $(d_\mathbf{k}^r,d_\mathbf{k}^{r\dagger})$  represent the annihilation and creation operators for Rob's particle and
antiparticle with $r=I,II$ in the distinct area, satisfying $\{c_\mathbf{k}^r,c_{\mathbf{k}'}^{r'\dagger}\}
=\{d_\mathbf{k}^r,d_\mathbf{k'}^{r'\dagger}\}=\delta(r-r')\delta(\mathbf{k}-\mathbf{k}')$.
The creation and annihilation operators for the Minkowski and Rindler's coordinates are connected by the Bogoliubov transform \cite{Barnett2002},
\begin{eqnarray}\label{eq}
	a_\mathbf{k}=\cos \vartheta c_\mathbf{k}^I-\sin \vartheta d_{-\mathbf{k}}^{II\dagger},
	b_{-\mathbf{k}}^\dagger=\sin \vartheta c_\mathbf{k}^I+\cos \vartheta d_{-\mathbf{k}}^{II\dagger},
\end{eqnarray}
with $\cos \vartheta=(\exp(-2\pi*\omega*c/a)+1)^{-1/2}$, $\omega$ is the frequency of the
Dirac particle and $c$ is the speed of light in vacuum.
Here, we employ a single-mode approximation, which is valid under the assumption that Rob's detector is sensitive to a single-particle mode in region $I$. This assumption allows us to approximate the frequency $\omega_A$ observed
by Alice to as being equivalent to the frequency $\omega_R$ observed by Rob, such that $\omega_A\sim\omega_R=\omega$. However, it is important to note that in general scenarios, the validity of the single-mode approximation may be compromised \cite{Alsing2006}.

We consider that Alice and Rob share the following initial state,
\begin{eqnarray}\label{state}
	|\Phi\rangle_{AR}=\cos\varphi|0\rangle_A|0\rangle_R+
	\sin\varphi|1\rangle_A|1\rangle_R,
\end{eqnarray}
where $\varphi\in[0, \pi/2]$. In the Minkowski vacuum, Rob's particle is in a general entangled two modes squeezed state:
\begin{eqnarray}\label{s0}
	|0\rangle_R=\cos \vartheta|0_k\rangle_I|0_{-k}\rangle_{II}
	+\sin \vartheta|1_k\rangle_I|1_{-k}\rangle_{II},
\end{eqnarray}
where the parameter $\vartheta$ is the same as eq.(\ref{eq}).
The excited state is defined by $|1\rangle_R=|1_k\rangle_I|0_{-k}\rangle_{II}$.
Hence, in terms of Minkowski modes for Alice and the Rindler modes for Rob we have
\begin{eqnarray}
|\Phi\rangle_{A,I,II} =\cos\varphi\cos\vartheta|0\rangle_A|0\rangle_I|0\rangle_{II}
+\cos\varphi\sin\vartheta|0\rangle_A|1\rangle_I|1\rangle_{II}
+\sin\varphi|1\rangle_A|1\rangle_I|0\rangle_{II}.
\end{eqnarray}

{\bf Bell-Mermin-Klyshko inequalities}
Consider observables $X,X'$ and $Y,Y'$ in Hilbert spaces $H_X$ and $H_Y$,
respectively, $X=n_i\sigma_i$, $X'=n_i'\sigma_i$,
$Y=m_i\sigma_i$, $Y'=m_i'\sigma_i$, where $\sigma_i$ are the Pauli operators and
$n_i,n_i',m_i,m_i'$ are unit vectors in three-dimensional Euclidean space.
The eigenvalues of these operators are $\pm1$.
The Bell operator $B_2$ is defined as
\begin{eqnarray}
B_2=\frac12(X\otimes Y+X'\otimes Y+X\otimes Y'-X'\otimes Y').
\end{eqnarray}
For states that admit local hidden variable models, in which correlations are explained by a hypothetical classical common cause (the hidden variable) within the common past light-cone of the measurement events \cite{Bell1964}, one has
the Bell inequality $|\langle B_2\rangle|\le1$ \cite{Clauser1969}.
If the Bell inequality is violated, the state is nonlocally correlated \cite{Brunner2014}.
Extending the Bell inequality to multipartite
systems, one has the Bell-Mermin-Klyshko inequality given by the operator
\begin{eqnarray}
B_n=\frac12B_{n-1}(O_n+O_n')+\frac12B_{n-1}'(O_n-O_n'),
\end{eqnarray}
where $n=2,3,4,...,$ $B_1=O_1$ and $B_1'=O_1'$,
$O_i$ and $O_i'$ can be chosen as $O_i=n_i\sigma_i$ and $O_i'=n_i'\sigma_i$
with $n_i=(\sin\theta_i\cos\phi_i,\sin\theta_i\sin\phi_i,\cos\theta_i)$ and
$n_i'=(\sin\theta_i'\cos\phi_i',\sin\theta_i'\sin\phi_i',\cos\theta_i')$.
When $O_n=\pm O_n'$ the Bell-Mermin-Klyshko inequality is given by
\begin{eqnarray}
|\langle B_n\rangle|\le1
\end{eqnarray}
for locally correlated states. Quantum mechanically one has
\begin{eqnarray}
|\langle B_n\rangle|\le 2^{\frac{n-1}{2}}.
\end{eqnarray}
In this work, we specifically focus on the
tripartite case. The Bell-Mermin-Klyshko (BMK3) operator has the form,
\begin{eqnarray}\label{bmk3}
B_3=\frac12(O_1O_2O_3'+O_1O_2'O_3+O_1'O_2O_3-O_1'O_2'O_3').
\end{eqnarray}

{\bf One-way information deficit}
For a bipartite state $\rho_{AB}$ in $\mathbbm{C}^2\otimes \mathbbm{C}^2$,
the one-way information deficit (OID) is given by \cite{Wang2013},
\begin{equation}\label{oqd}
\mathrm{OID}=\min_{\Pi_i} S(\rho_{AB}')-S(\rho_{AB}),
\end{equation}
where $\rho_{AB}'=\Sigma_i\Pi_i \rho_{AB}\Pi_i$,
$\Pi_i$ is a projector on subsystem $A$ satisfying $\Pi_i\Pi_j
=\delta_{ij}\Pi_i$ and $\Sigma_i\Pi_i=\mathbbm{1}$, $S(\rho)=-{\rm Tr}[\rho\log\rho]$
is the von Neumann entropy.
We denote $\mathrm{ID}=S(\rho_{AB}')-S(\rho_{AB})$ the information deficit (ID).
OID is equivalent to the minimal relative entropy distance between
the set of post-measurement states $\rho_{AB}'$ and the initial state $\rho_{AB}$.

\section{Main results}
We calculate in this section the mean value $\langle B_3\rangle$ of the BMK3 operator, the ID
and OID for the state $|\Phi\rangle_{A,I,II}$.

\begin{figure}[htbp!]
\centering
\includegraphics[width=.48\textwidth]{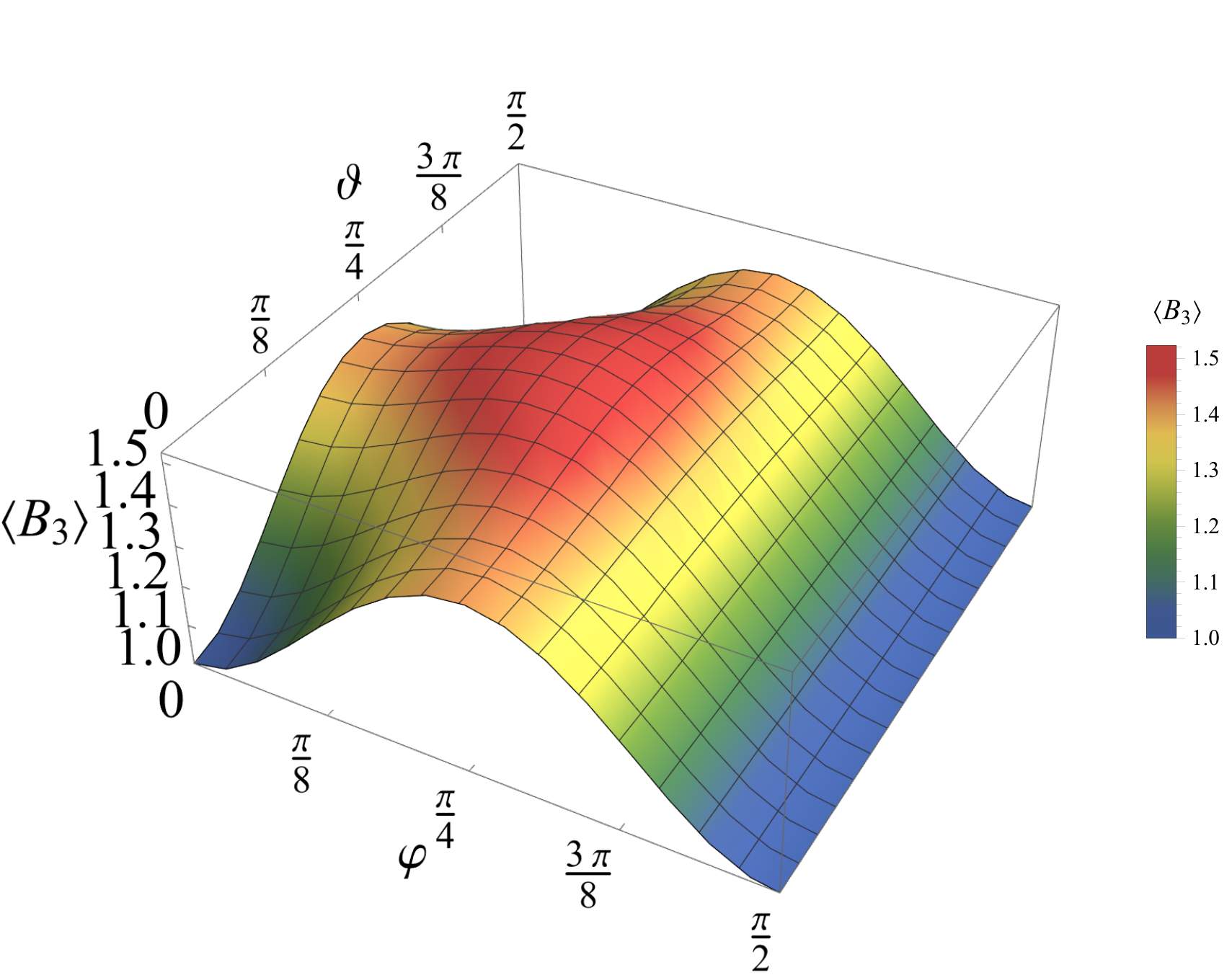}
\caption{Mean value of the Bell-Mermin-Klyshko operator with respect to $\varphi$ and $\vartheta$ for $|\Phi\rangle_{A,I,II}$.
}
\label{f1}
\end{figure}

Fig. \ref{f1} illustrates the mean value $\langle B_3\rangle$ with respect to
the $\varphi$ and $\vartheta$ for $|\Phi\rangle_{A,I,II}$, where
the parameters $\theta_1, \theta_2, \theta_3, \theta_1',\theta_2',\theta_3',
\phi_1,\phi_2,\phi_3,\phi_1',\phi_2'$ and $\phi_3'$ have been optimally maximized.
The BMK3 in equality $\langle B_3\rangle\leq 1$ is violated for most states, except
for the product states $|1\rangle_A|1\rangle_I|0\rangle_{II}$ ($\varphi=\pi/2$), $|0\rangle_A|0\rangle_I|0\rangle_{II}$ ($(\varphi,\vartheta)=(0,0)$) and $|0\rangle_A|1\rangle_I|1\rangle_{II}$ ($(\varphi,\vartheta)=(0,\pi/2)$). All the rest states violate the BMK3 inequality which have quantum nonlocal properties. It can be observed that $\langle B_3\rangle$ is
symmetric about $\vartheta=\pi/4$, which corresponds to Rob's acceleration $a=\pm\infty$.
As $a$ changes from $\pm\infty$ to 0, the violation of the BMK3 inequality decreases.
For fixed $\varphi=0$, $\langle B_3\rangle$ increases with $\vartheta$ initially, and reaches a maximum at $\vartheta=\pi/4$, and then decreases to a minimum at
$\vartheta=0$ or $\pi/2$. This indicates that quantum nonlocality diminishes as the magnitude of acceleration $|a|$ decreases. However, for $\varphi=\pi/2$, acceleration does not affect nonlocality.

For fixed $\vartheta=0$, $\langle B_3\rangle$ initially increases and then decreases with $\varphi$. However, for $\vartheta=\pi/4$, $\langle B_3\rangle$ first increases 
slowly and then descends swiftly. This suggests that different accelerations lead to  distinct changes in nonlocality. The BMK3 inequality is violated for most states, except at two endpoints. 

\begin{figure}[htbp!]
\centering
\includegraphics[width=.48\textwidth]{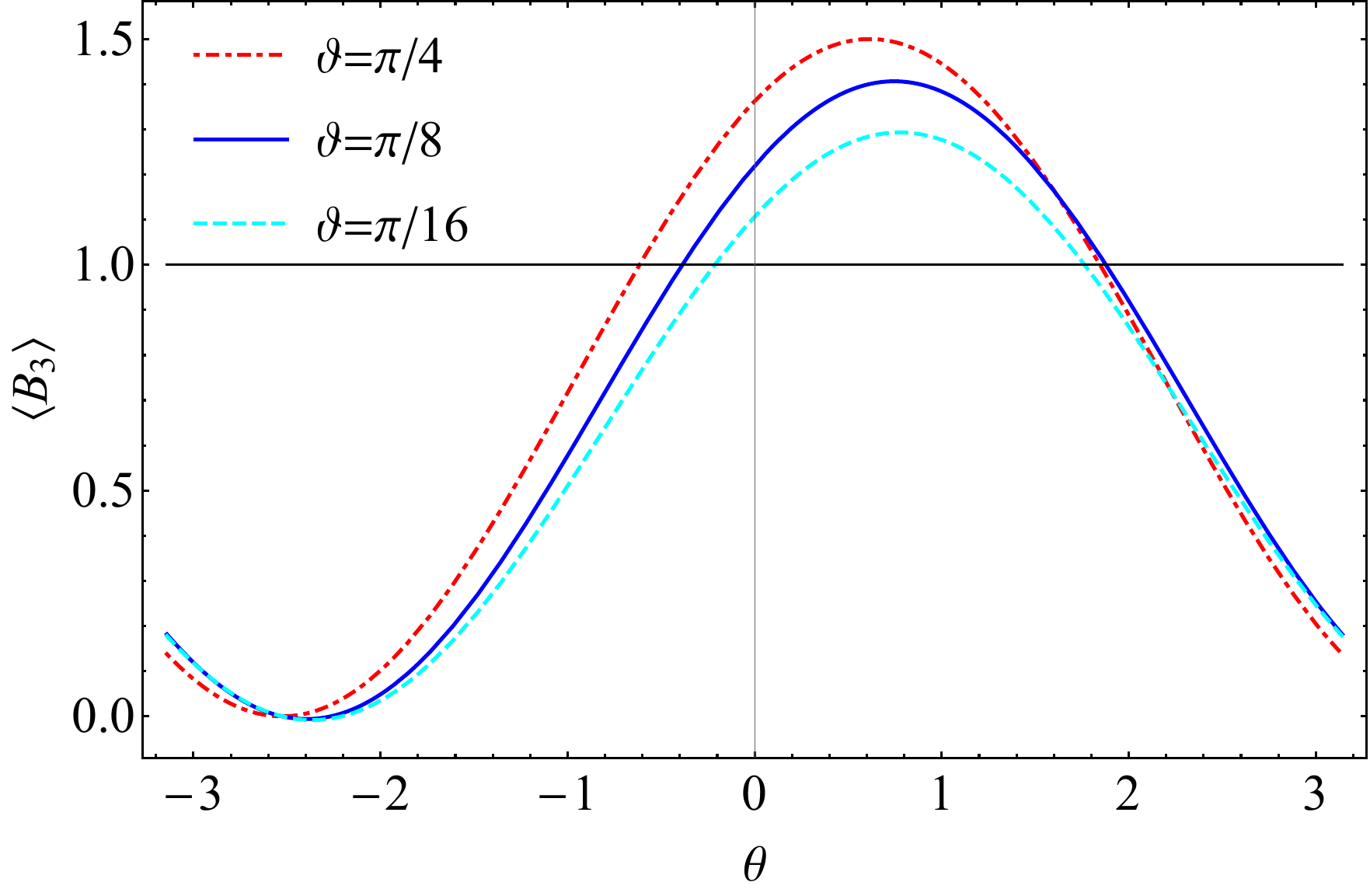}
\caption{Bell-Mermin-Klyshko inequality with respect to $\theta$ for $\varphi=\pi/4$
and $\vartheta=\pi/4, \pi/8, \pi/16$ (represented by
red dot-dashed, blue solid and cyan dashed lines, respectively).
}
\label{f2}
\end{figure}

By setting $\theta_2=\theta$ and
taking $\theta_1=0,\theta_3=-0.61548,\theta_1'=1.5708,\theta_2'=-0.61548,\theta_3'=0.61548,
\phi_1=-0.35236,\phi_2=0.268617,\phi_3=-0.268617,
\phi_1'=-0.268617,\phi_2'=0.268617,\phi_3'=-0.268617$,
we plot $\langle B_3\rangle$
as a function of $\theta$
in Fig. \ref{f2}, where the red dot-dashed, blue solid and cyan dashed
lines are for $\vartheta=\pi/4$, $\pi/8$ and $\pi/16$, respectively.
The behavior of $\langle B_3\rangle$ with respect to $\theta$
follows a decreasing trend initially,
then increases, and eventually decreases
again. Only in a small region of the parameter $\theta$ the BMK3 inequality is violated.
From Fig. \ref{f2}, it can be observed that the range
of BMK3 violation shrinks from $\pi/4$ to $\pi/8$ and then
to $\pi/16$, indicating that as the
acceleration increases, the violation zones become smaller.

Now we proceed to investigate the one-way information deficit.
The following results demonstrate the impact of acceleration on the OID.
If we trace out region II of the state, we
obtain the following expression,
\begin{eqnarray}\label{ai}	
\rho_{A,I}=
\left(
\begin{array}{cccc}
 \cos ^2(\varphi ) \cos ^2(\vartheta ) & 0 & 0 & \sin (\varphi ) \cos (\varphi ) \cos (\vartheta ) \\
 0 & \cos ^2(\varphi ) \sin ^2(\vartheta ) & 0 & 0 \\
 0 & 0 & 0 & 0 \\
 \sin (\varphi ) \cos (\varphi ) \cos (\vartheta ) & 0 & 0 & \sin ^2(\varphi ) \\
\end{array}
\right).
\end{eqnarray}
We assess the projective measurement on subsystem $A$ defined by
\begin{eqnarray}
	\Pi_0=\frac{I_2+\vec{n}\cdot \vec{\sigma}}{2}\otimes I_2,~~~
	\Pi_1=\frac{I_2-\vec{n}\cdot \vec{\sigma}}{2}\otimes I_2,
\end{eqnarray}
where $\vec{n}=(\sin\eta\cos\xi,\sin\eta\sin\xi,\cos\eta)$ and $I_2$ denotes the $2\times 2$
identity matrix. In order to compute the OID, we optimize over the angles $\eta$ and $\xi$. It has been verified that the optimal measurement angle is dependent solely on the parameter $\eta$, with $\xi$ set to 0. For further 
details regarding the calculation of the OID, refer to \cite{Ye2016}.

\begin{figure}[htbp!]
\centering
\subfigure[$\varphi=\pi/4$]{
\includegraphics[width=.38\textwidth]{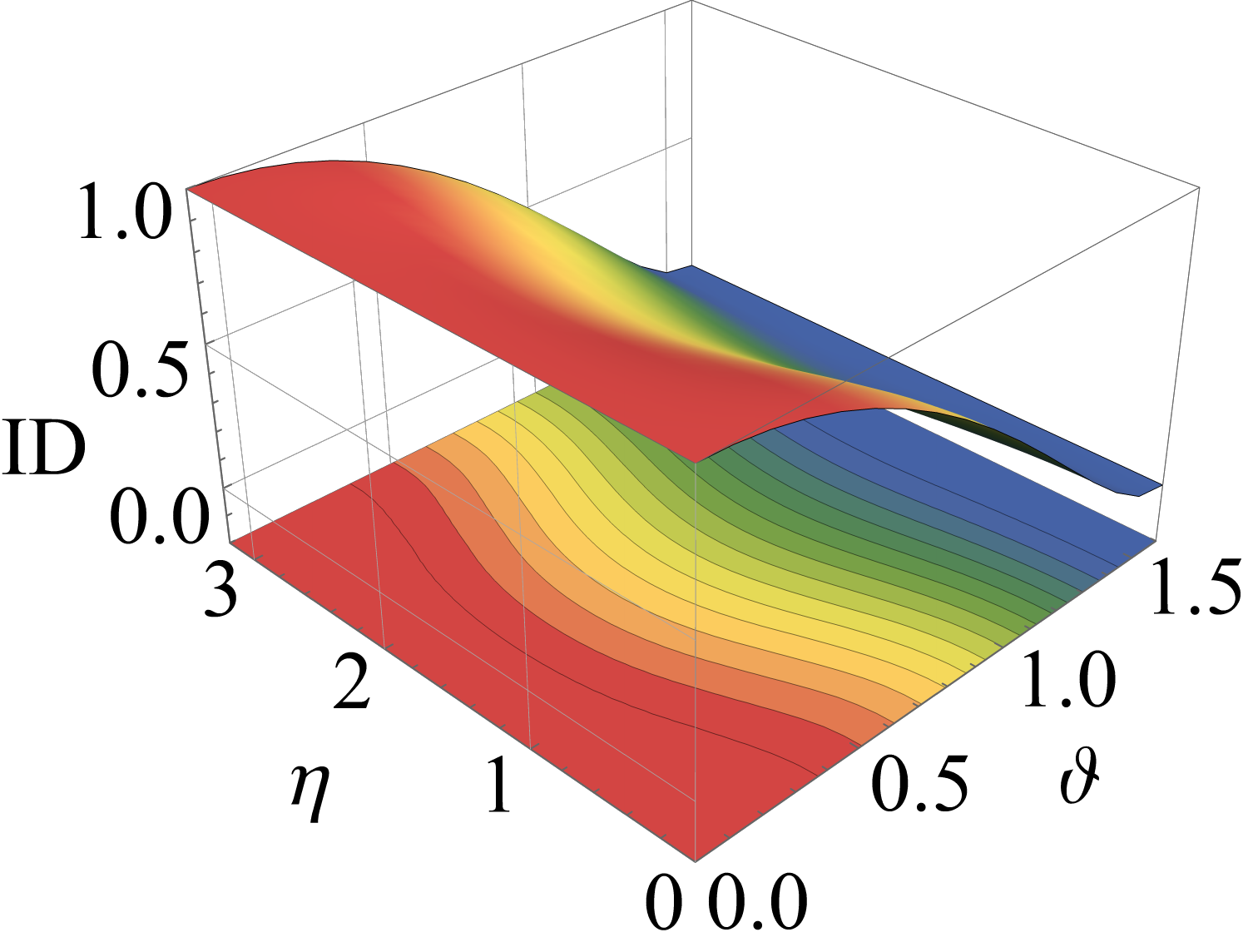}
}
\subfigure[$\varphi=\pi/8$]{
\includegraphics[width=.38\textwidth]{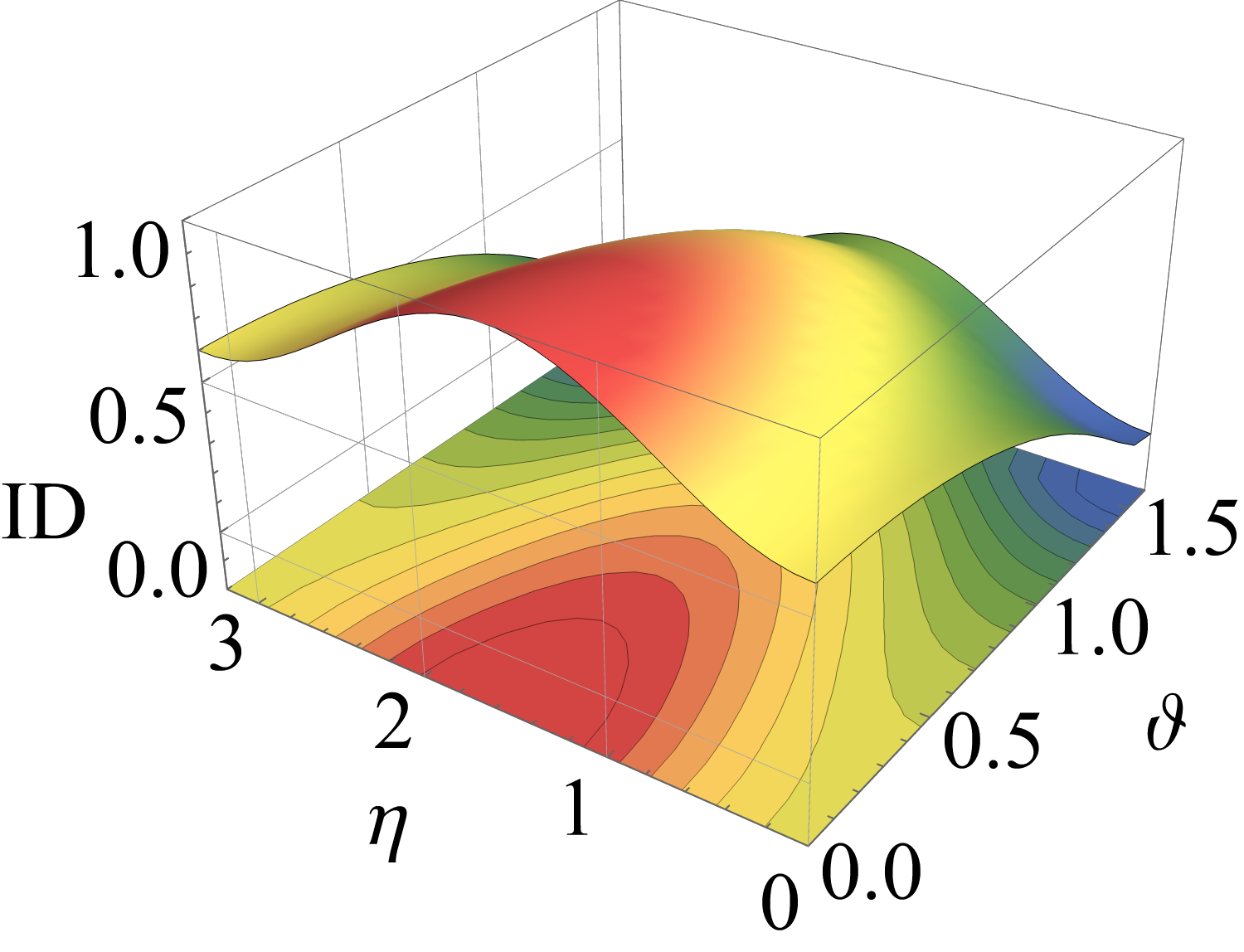}
}
\caption{ID of $\rho_{A,I}$ with respect to $\vartheta$ and $\eta$
for $\varphi=\pi/4$ and $\varphi=\pi/8$. From the figures, we observe that the minimum
information deficit is attained at $\eta=\pi/2$ for (a) and $\eta=0, \pi$ for (b).
}
\label{f3}
\end{figure}

Figure \ref{f3} (a) illustrates the case when $\varphi=\pi/4$.
The information deficit varies with $\vartheta$ and the angle $\eta$.
We observe that the information deficit decreases as
$\vartheta$ increases. The contour graphics below depicts the profile
of the information deficit. The information deficit is symmetric about $\eta=\pi/2$. However, it reaches the minimum value at $\eta=\pi/2$. The corresponding optimal measurement bases are $(|0\rangle\pm|1\rangle)/\sqrt{2}$. Thus, we derive the analytical expression
for OID below. From Figure 3 (a), we see that the optimal measurement angle
for $\rho_{A,I}'$ is $\eta=\pi/2$. Therefore, we have
\begin{eqnarray}
\lambda_1'&=&\lambda_2'=\frac{1}{16} \left(4-\sqrt{2\cos (4 \vartheta )+14}\right),\cr
\lambda_3'&=&\lambda_4'=\frac{1}{16} \left(4+\sqrt{2\cos (4 \vartheta )+14}\right),
\end{eqnarray}
where $\lambda_i's$ are the optimal eigenvalues of $\rho_{A,I}'$.
Thus, we obtain the analytical expression for the one-way information deficit for $\rho_{A, I}$ as follows,
\begin{eqnarray} \label{17}
\mathrm{OID}&=&\kappa_1\log\kappa_1+\kappa_2\log\kappa_2-2(\lambda_1'\log\lambda_1'+\lambda_3'\log\lambda_3').
\end{eqnarray}
Here, $\kappa_1=(1-\cos(2\vartheta))/4$ and $\kappa_2=(3+\cos(2\vartheta))/4$.

In Figure \ref{f3} (b), $\varphi=\pi/8$. The information deficit (ID) is symmetric with
respect to $\eta=\pi/2$. It is observed that the
ID decreases as $\vartheta$ increases. The contour in the figure
also illustrates the trend of the ID. The optimal measurement angle
is attained at $\eta=0$ or $\pi$. The corresponding optimal measurement bases are $|0\rangle$ or $|1\rangle$. We have the following analytical expression for the
one-way information deficit,
\begin{eqnarray} \label{18}
\mathrm{OID}=&-&[\cos ^2\left(\frac{\pi }{8}\right) \cos ^2(\vartheta )]\log[\cos ^2\left(\frac{\pi }{8}\right) \cos ^2(\vartheta )]
-[\sin ^2\left(\frac{\pi }{8}\right)]\log[\sin ^2\left(\frac{\pi }{8}\right)]\nonumber\\
&+&[\cos ^2\left(\frac{\pi }{8}\right) \cos ^2(\vartheta )+\sin ^2\left(\frac{\pi }{8}\right)]\log[\cos ^2\left(\frac{\pi }{8}\right) \cos ^2(\vartheta )+\sin ^2\left(\frac{\pi }{8}\right)].
\end{eqnarray}

\begin{figure}[htbp!]
\centering
\includegraphics[width=.48\textwidth]{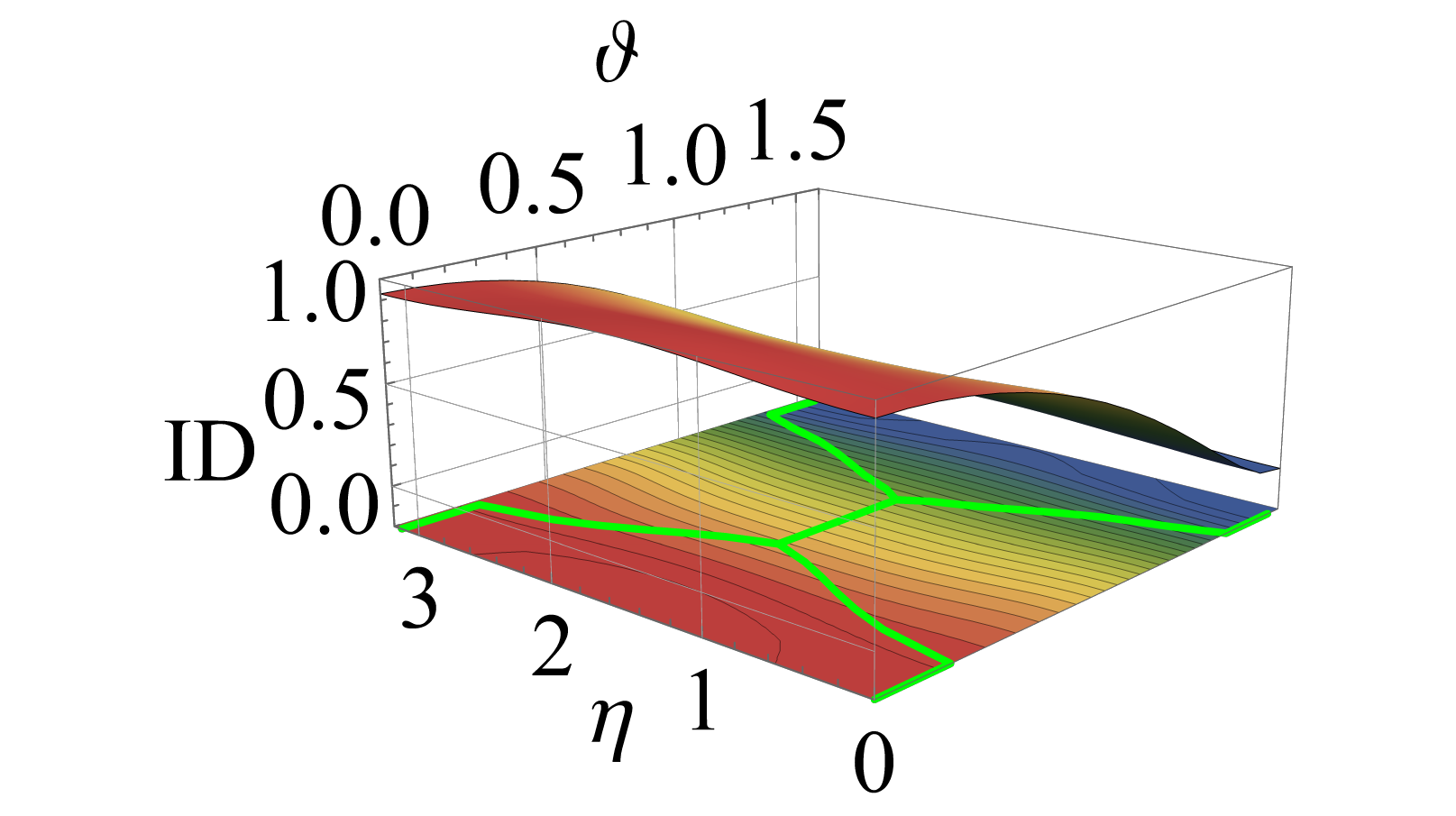}
\caption{ID of $\rho_{A,I}$ with respect to $\vartheta$ and $\eta$
for $\varphi=\pi/5$. The figure shows the numerical derivation of the minimum
at $\eta$ (green line).
}
\label{f5}
\end{figure}

Different from Figure 3 (a) and (b), for $\varphi=\pi/5$ the optimal measurement angle $\eta$ is not fixed. The optimal value of $\eta$ alternates between the
regions of $(0,\pi/2)$ and $(\pi/2,\pi)$, as observed
from the green lines in Figure \ref{f5}. The ID is symmetric with respect to the
angle $\eta$ at $\pi/2$. For $\eta=0$ and $\pi$ the ID is minimal at some
regions. Moreover, there is also a small
region around $\eta=\pi/2$ such that the optimal value is attained.
Thus, the analytical expression of the
one-way information deficit for $\varphi=\pi/5$ is incomplete.

\begin{figure}[htbp!]
\centering
\subfigure[$\varphi=\pi/4$]{
\includegraphics[width=.38\textwidth]{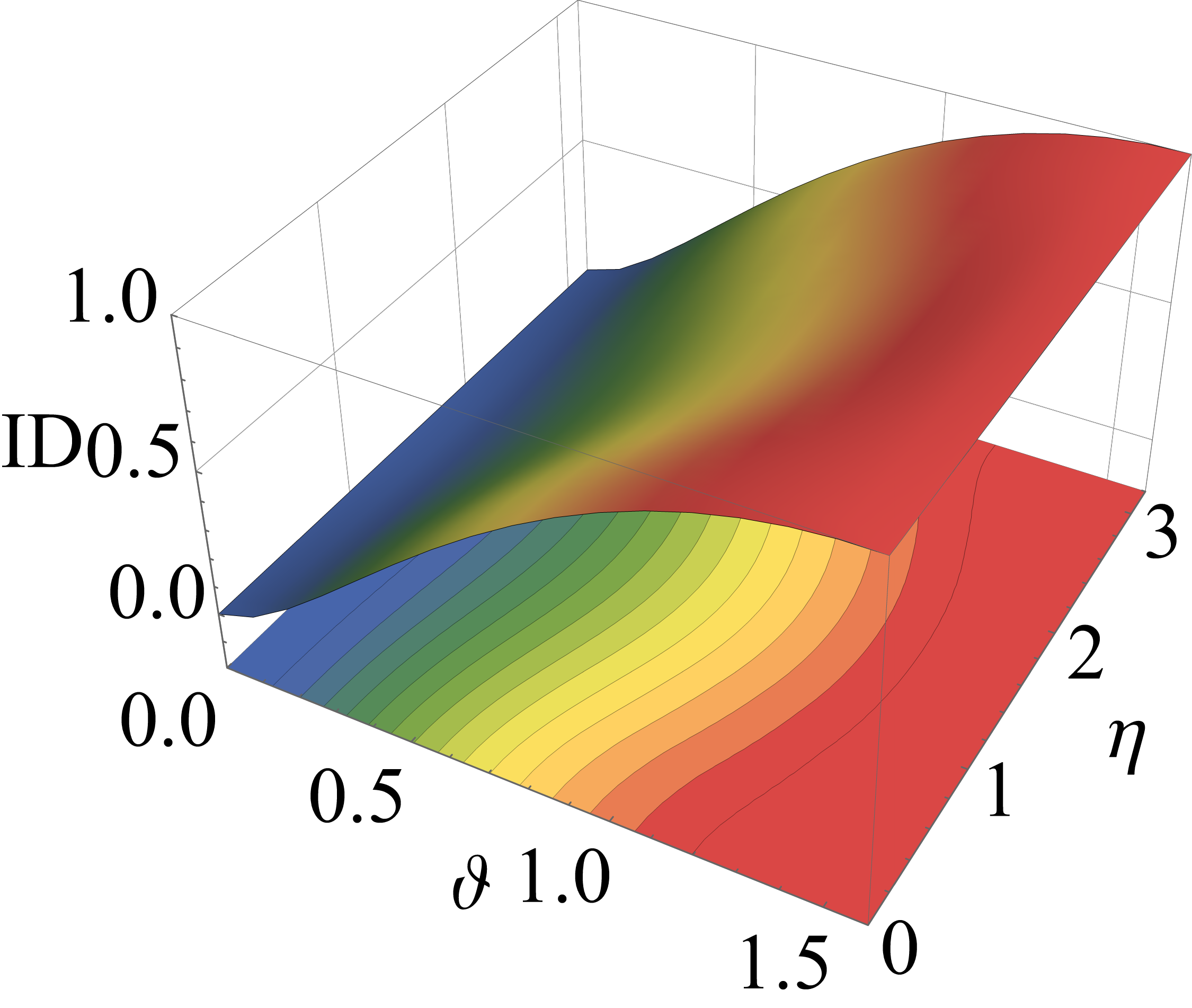}
}
\subfigure[$\varphi=\pi/8$]{
\includegraphics[width=.38\textwidth]{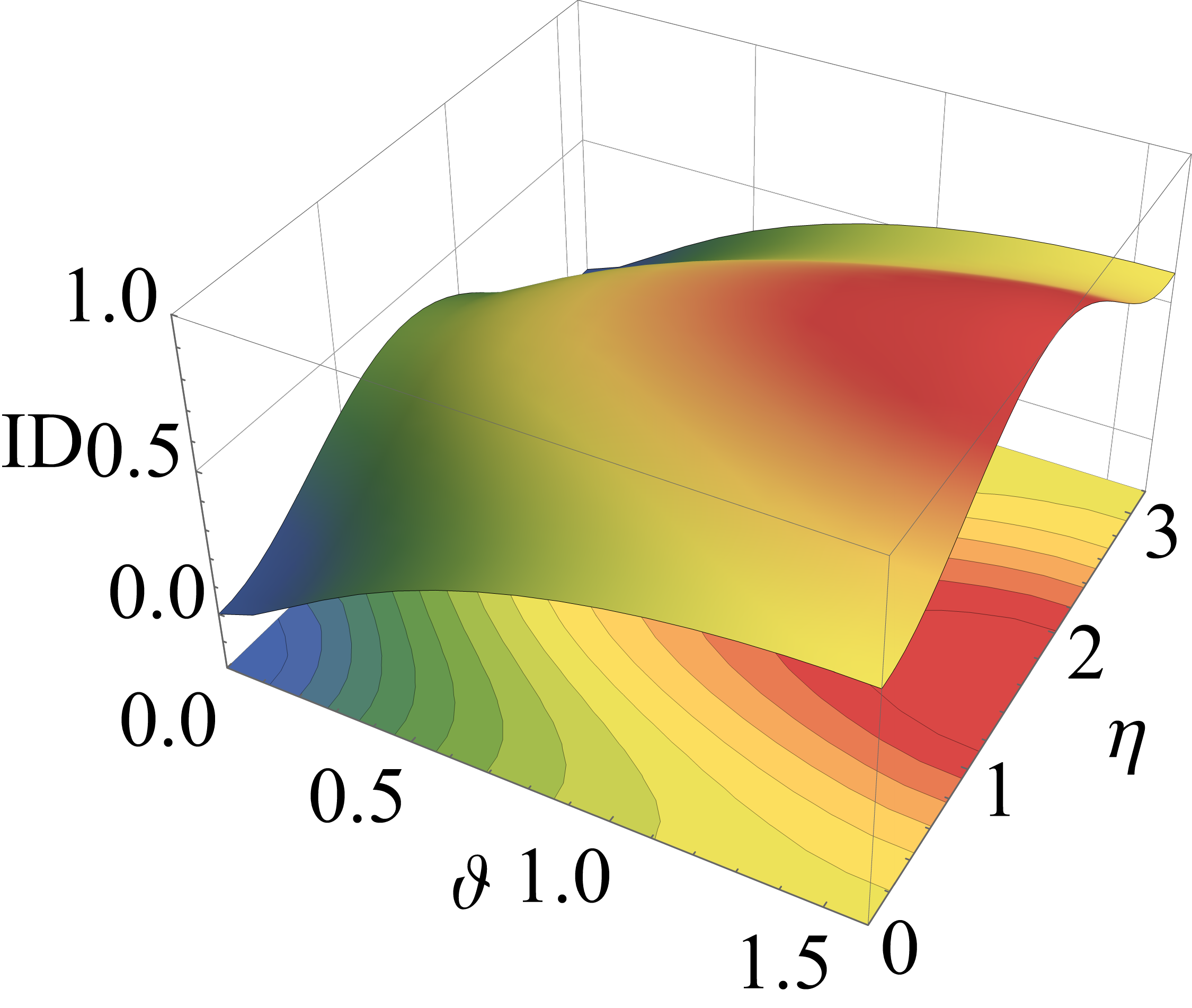}
}
\caption{ID of $\rho_{A,II}$ with respect to $\vartheta$ and $\eta$
for $\varphi=\pi/4$ and $\varphi=\pi/8$. The figure shows that the minimum ID is attained
at $\eta=\pi/2$ in Figure (a), and $\eta=0$ and $\pi$ in Figure (b).
}
\label{f7}
\end{figure}

If we trace out the system $I$, we get
\begin{eqnarray}\label{aii}
	\rho_{A,II}=
\left(
\begin{array}{cccc}
 \cos ^2(\varphi ) \cos ^2(\vartheta ) & 0 & 0 & 0 \\
 0 & \cos ^2(\varphi ) \sin ^2(\vartheta ) & \sin (\varphi ) \cos (\varphi ) \sin (\vartheta ) & 0 \\
 0 & \sin (\varphi ) \cos (\varphi ) \sin (\vartheta ) & \sin ^2(\varphi ) & 0 \\
 0 & 0 & 0 & 0 \\
\end{array}
\right).
\end{eqnarray}
Similar to the state $\rho_{A,I}$, we measure the system $A$. Figure \ref{f7} shows that the ID varies with the parameters $\vartheta$ and $\varphi$.
In Figure \ref{f7}(a) with $\varphi=\pi/4$, the ID is symmetric about $\eta=\pi/2$.
However, it is different from Figure \ref{f3}, as the ID increases as $\vartheta$ grows.
We obtain the minimal ID at the optimal angle $\eta=\pi/2$. The contour
figure also shows that the minimum occurs at this point.
For the state $\rho_{A,II}$, the ID changes with $\vartheta$
and $\eta$, and we find the minimum at $\eta=\pi/2$.
The corresponding optimal measurement bases are $(|0\rangle\pm|1\rangle)/\sqrt{2}$. The analytical expression for the ID is given by
\begin{eqnarray} \label{20}
\mathrm{OID}=&-&2(\mu_1\log\mu_1+\mu_2\log\mu_2)+
[\frac{1}{4}(1+\cos (2\vartheta ))]\log[\frac{1}{4}(1+\cos (2\vartheta ))]\nonumber\\
&+&[\frac{1}{4} (3-\cos (2 \vartheta ))]\log[\frac{1}{4} (3-\cos (2 \vartheta ))]
,
\end{eqnarray}
where
$\mu_1=\frac{1}{16} \left(4- \sqrt{2\cos (4 \vartheta )+14}\right)$ and
$\mu_2=\frac{1}{16} \left(4+ \sqrt{2\cos (4 \vartheta )+14}\right)$.

Figure 5 (b) shows the case of $\varphi=\pi/8$. The ID also
grows as $\vartheta$ increases. The symmetric axis is at $\eta=\pi/2$.
For $\vartheta=0$, the ID changes with respect to $\eta$, first increasing
and then decreasing slowly. The optimal angle
is at the boundary $\eta=0, \pi$. The corresponding optimal measurement bases are $|0\rangle$ or $|1\rangle$. The analytical expression for one-way information
deficit is given by
\begin{eqnarray}\label{21}
\mathrm{OID}=&-&[\sin ^2\left(\frac{\pi }{8}\right)]\log[\sin ^2\left(\frac{\pi }{8}\right)]
-[\cos ^2\left(\frac{\pi }{8}\right) \sin ^2(\vartheta )]\log[\cos ^2\left(\frac{\pi }{8}\right) \sin ^2(\vartheta )]\nonumber\\
&+&[\sin ^2\left(\frac{\pi }{8}\right)+\cos ^2\left(\frac{\pi }{8}\right) \sin ^2(\vartheta )]\log[\sin ^2\left(\frac{\pi }{8}\right)+\cos ^2\left(\frac{\pi }{8}\right) \sin ^2(\vartheta )].
\end{eqnarray}

\begin{figure}[htbp!]
\centering
\includegraphics[width=.48\textwidth]{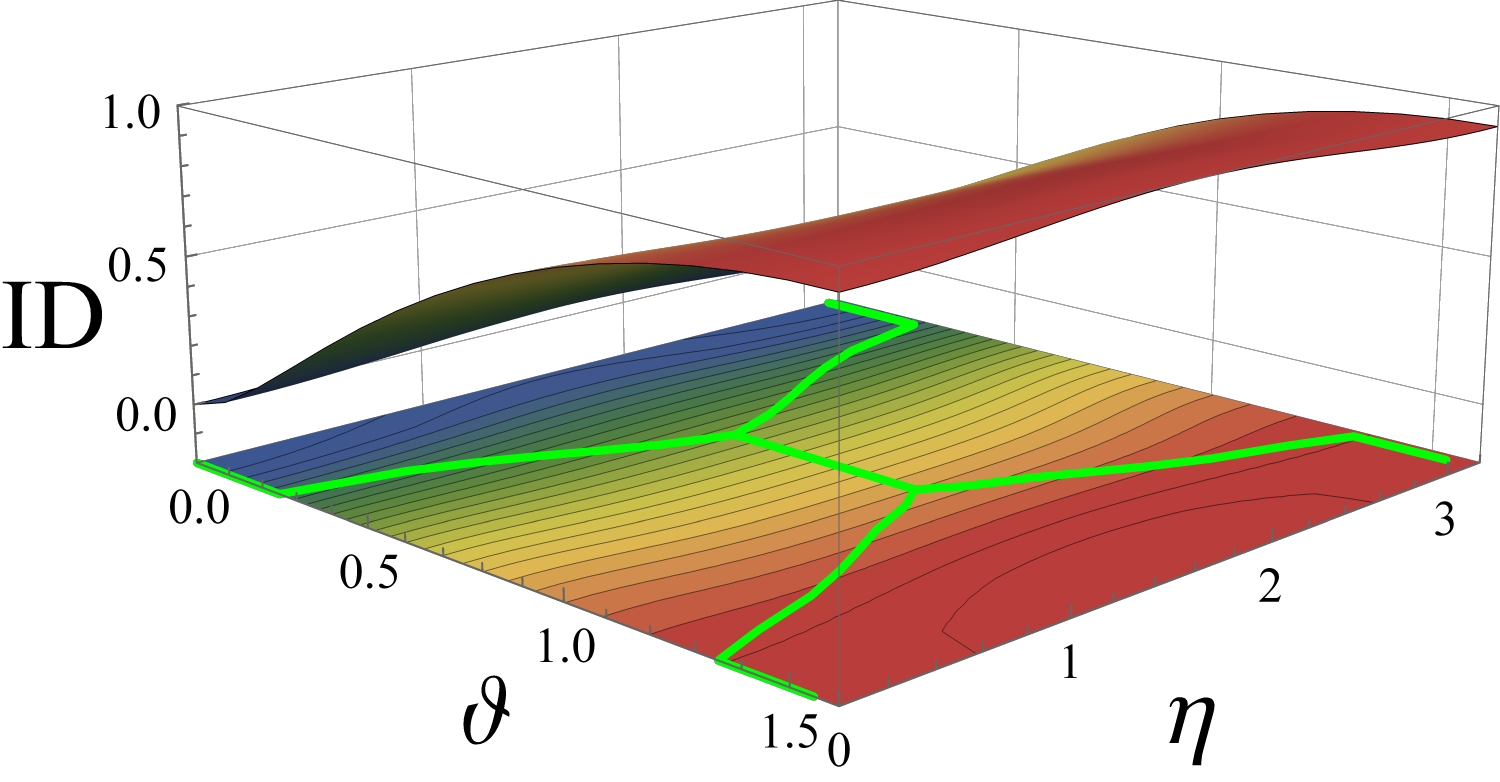}
\caption{ID with respect to $\vartheta$ and $\eta$
for $\varphi=\pi/5$ for $\rho_{A,II}$. The minimum ID occurs at the green line.
}
\label{f9}
\end{figure}

We also plot the case of $\varphi=\pi/5$ for the state $\rho_{A,II}$. The optimal angle $\eta$
with respect to $\rho_{A,II}$ is not a fixed value.
One can find the ID and obtain the optimal values at $\eta=0, \pi/2, \pi$ for
some $\vartheta$. However, for certain regions
between $(0,\pi/2)$ and $(\pi/2, \pi)$, only numerical results
can be derived. As $\vartheta$ increases, the ID also increases.
The contour of ID is shown in Figure \ref{f9}. The green lines
represent the optimal angle derived for the ID. The analytical expression
for the optimal angle is not completely obtained.

\begin{figure}[htbp!]
\centering
\subfigure[$\varphi=\pi/4$]{
\includegraphics[width=.28\textwidth]{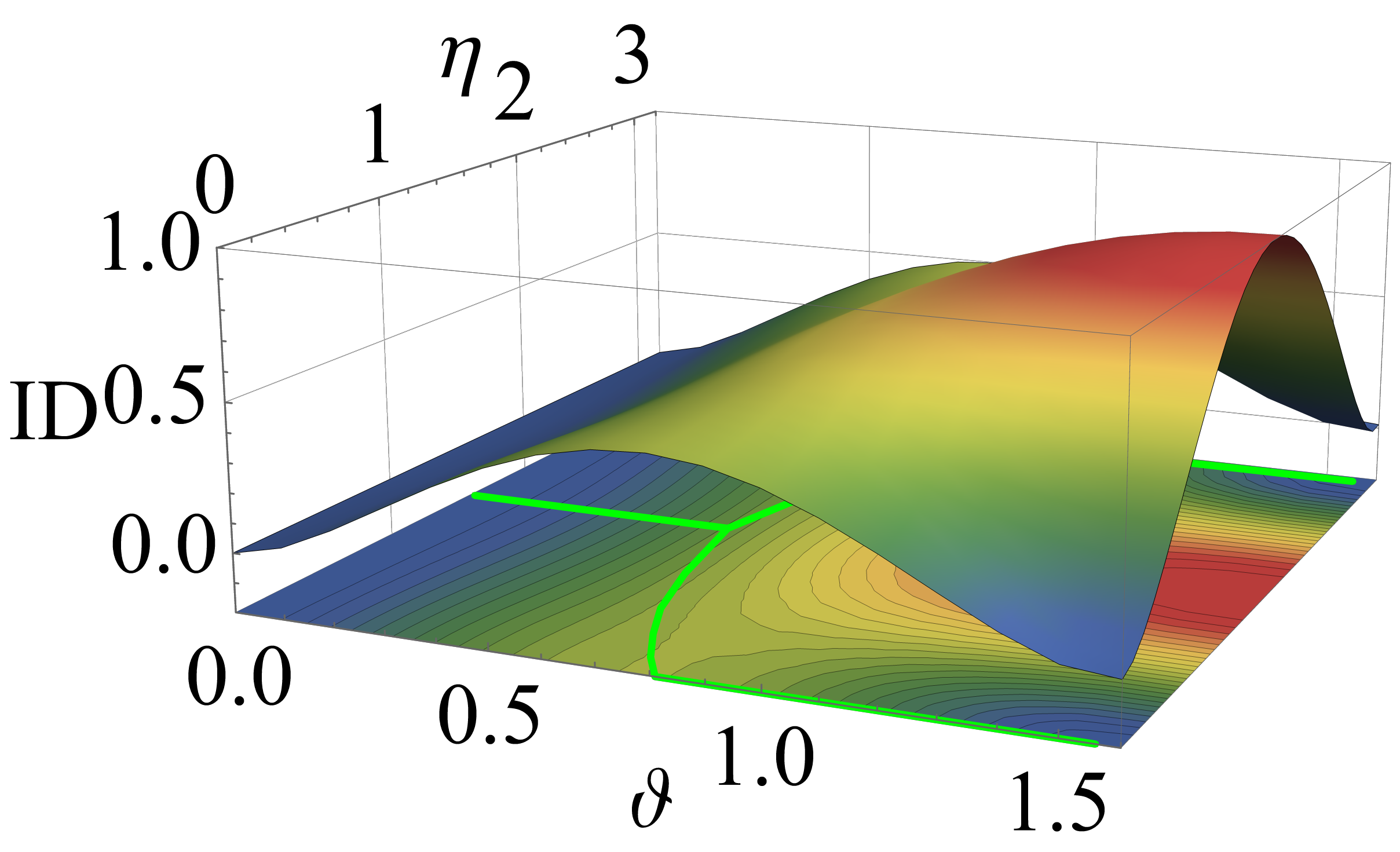}
}
\subfigure[$\varphi=\pi/5$]{
\includegraphics[width=.28\textwidth]{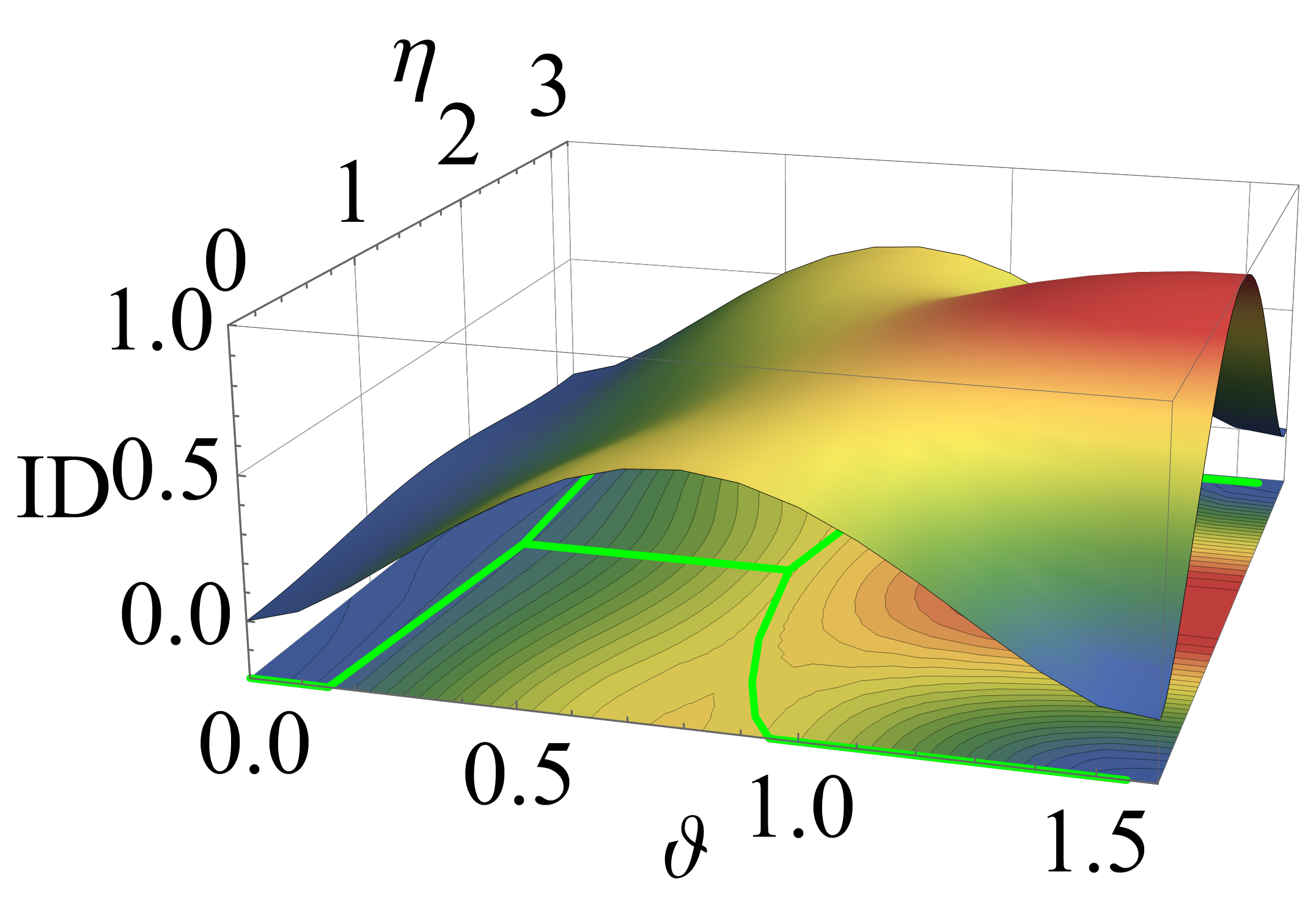}
}
\subfigure[$\varphi=\pi/8$]{
\includegraphics[width=.28\textwidth]{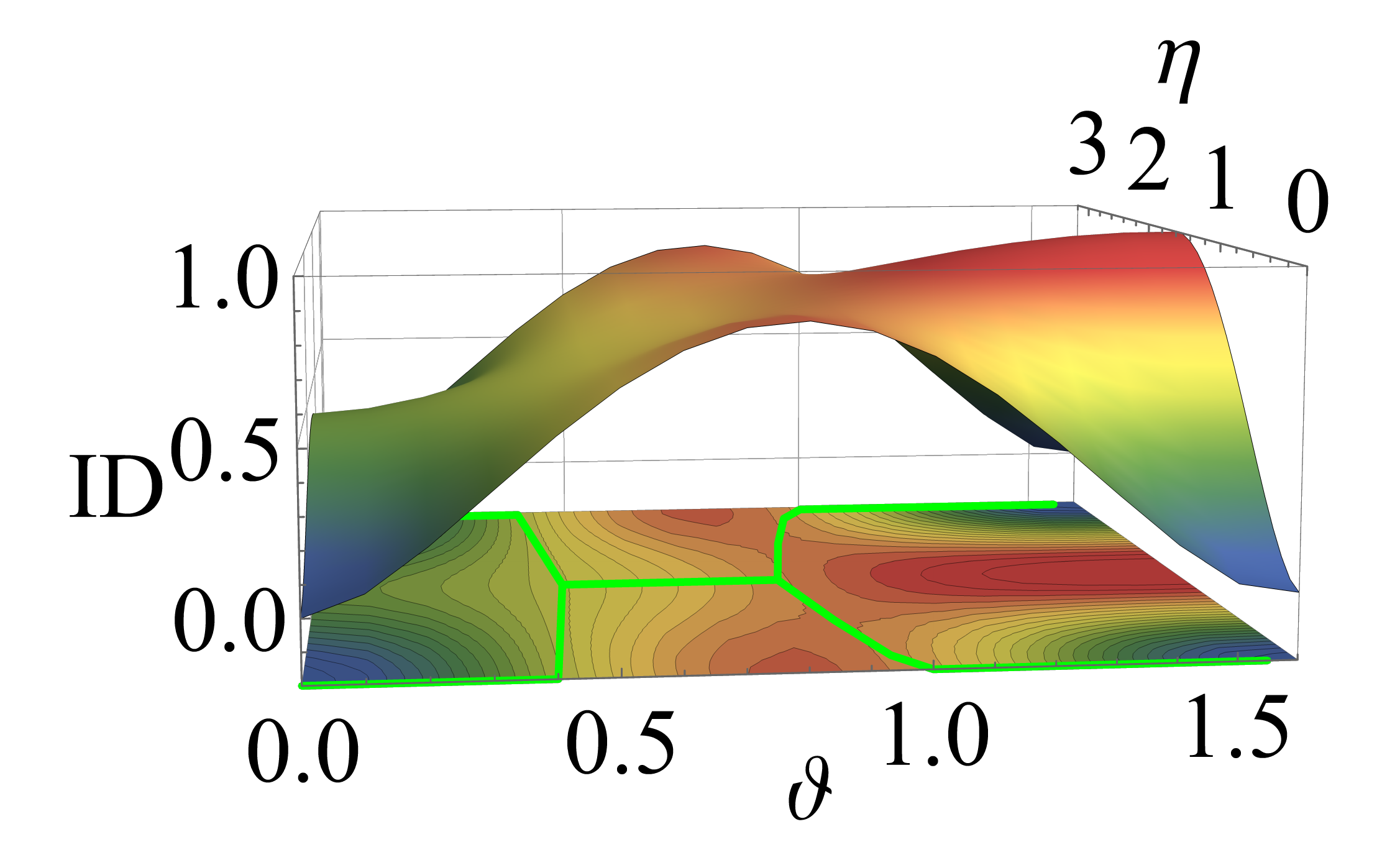}
}
\caption{ID of $\rho_{I,II}$ with respect to $\vartheta$ and $\eta$
for $\varphi=\pi/4$, $\varphi=\pi/5$ and $\varphi=\pi/8$.
The green line represents the minimal ID at $\eta$.
}
\label{iii}
\end{figure}

Now we consider the state by tracing over the system $A$,
\begin{eqnarray}
	\rho_{I,II}=
\left(
\begin{array}{cccc}
 \cos ^2(\varphi ) \cos ^2(\vartheta ) & 0 & 0 & \cos ^2(\varphi ) \sin (\vartheta ) \cos (\vartheta ) \\
 0 & 0 & 0 & 0 \\
 0 & 0 & \sin ^2(\varphi ) & 0 \\
 \cos ^2(\varphi ) \sin (\vartheta ) \cos (\vartheta ) & 0 & 0 & \cos ^2(\varphi ) \sin ^2(\vartheta ) \\
\end{array}
\right).
\end{eqnarray}
We analyze the state $\rho_{I,II}$ by performing measurements
on subsystem $I$. Figure \ref{iii} shows the ID with respect to
$\vartheta$ and $\eta$ for different values $\varphi=\pi/4$,
$\pi/5$ and $\pi/8$ corresponding to (a), (b) and (c), respectively.
For (a), the optimal ID values are found in some regions of $\eta$ (specifically
$\eta=0$, $\pi/2$ and $\pi$) alongside center optimal values.
For (b), optimal ID values are only found for
certain regions of $\vartheta$, while $\eta=0$, $\pi/2$ and $\pi$ remain optimal. 
The case for (c) is the same as for (b).
Some regions exhibit optimal minimal values at $(0,\pi/2)$ and $(\pi/2, \pi)$.
All three subfigures focus on $\eta=\pi/2$.
For $\eta=0$, the ID initially increases, then gradually decreases
as the $\vartheta$ grows. However, when $\eta=\pi/2$, the
ID continually increases until reaching its maximum as the $\vartheta$ increases.
Some optimal values occur at $\eta=\pi/2$. The contour of the green lines in all
subfigures represents the optimal values attained.
For $\varphi=\pi/4$, $\varphi=\pi/5$, $\varphi=\pi/8$ and $\rho_{I,II}$, the optimal
measurement angle $\eta$ is not a fixed value. It can be inferred from the figure
that $\eta=0,\pi/2$ for various values, but within the region $(0, \pi/2)$.
Thus, a completely analytical expression for the one-way information
deficit is not established.

\begin{figure}[htbp!]
\centering
\subfigure[$\rho_{A,I}$]{
\includegraphics[width=.31\textwidth]{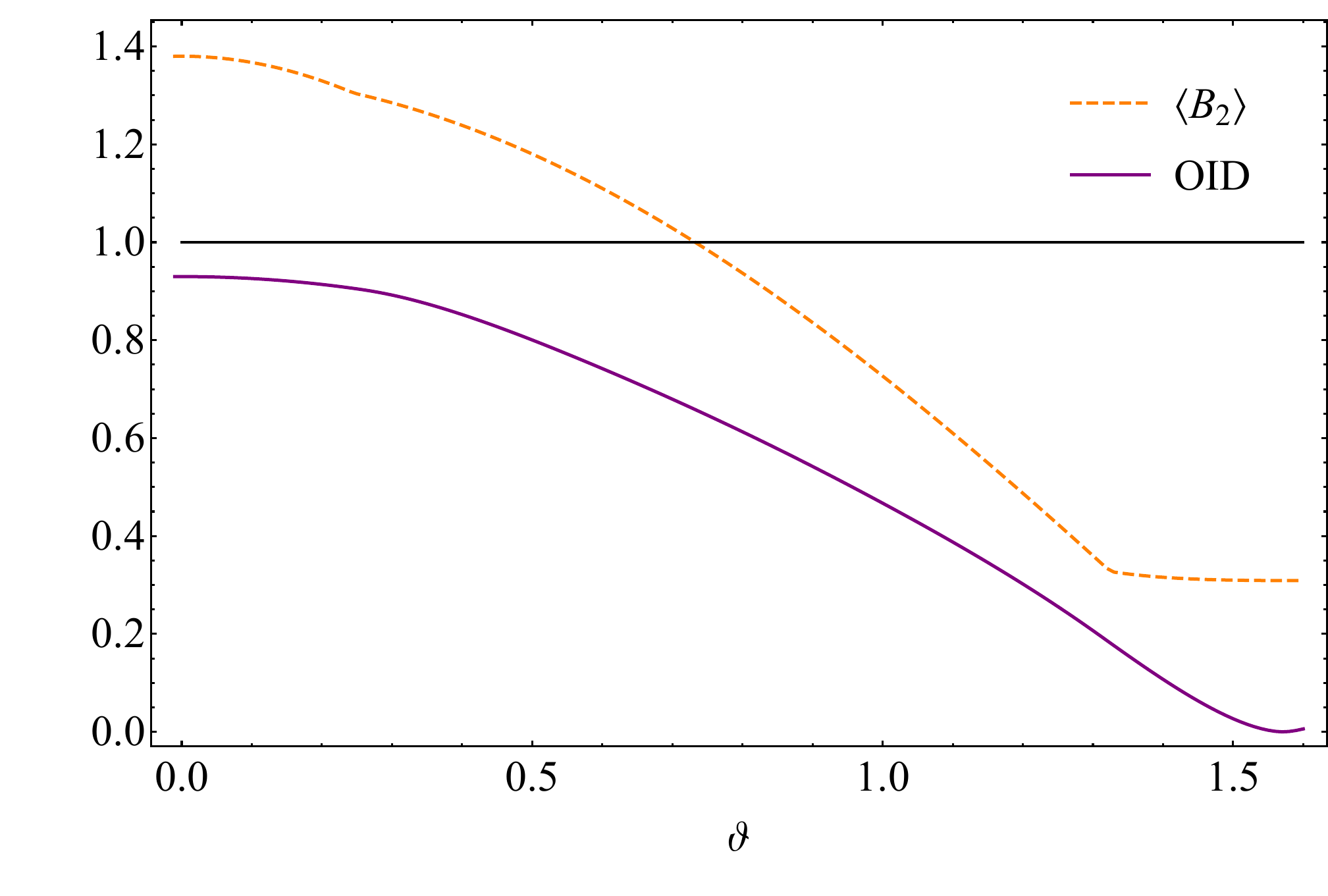}
}
\subfigure[$\rho_{A,II}$]{
\includegraphics[width=.31\textwidth]{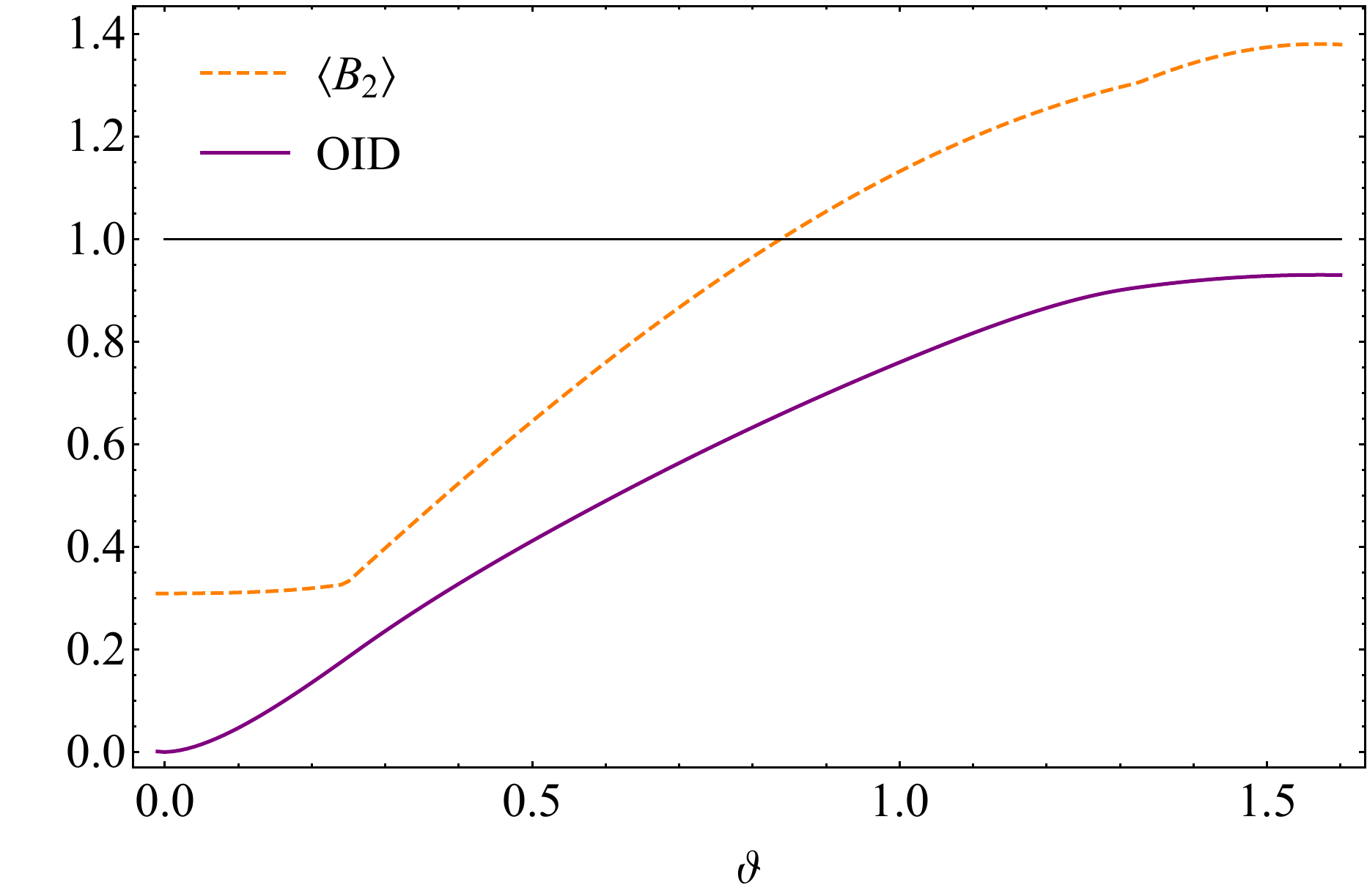}
}
\subfigure[$\rho_{I,II}$]{
\includegraphics[width=.31\textwidth]{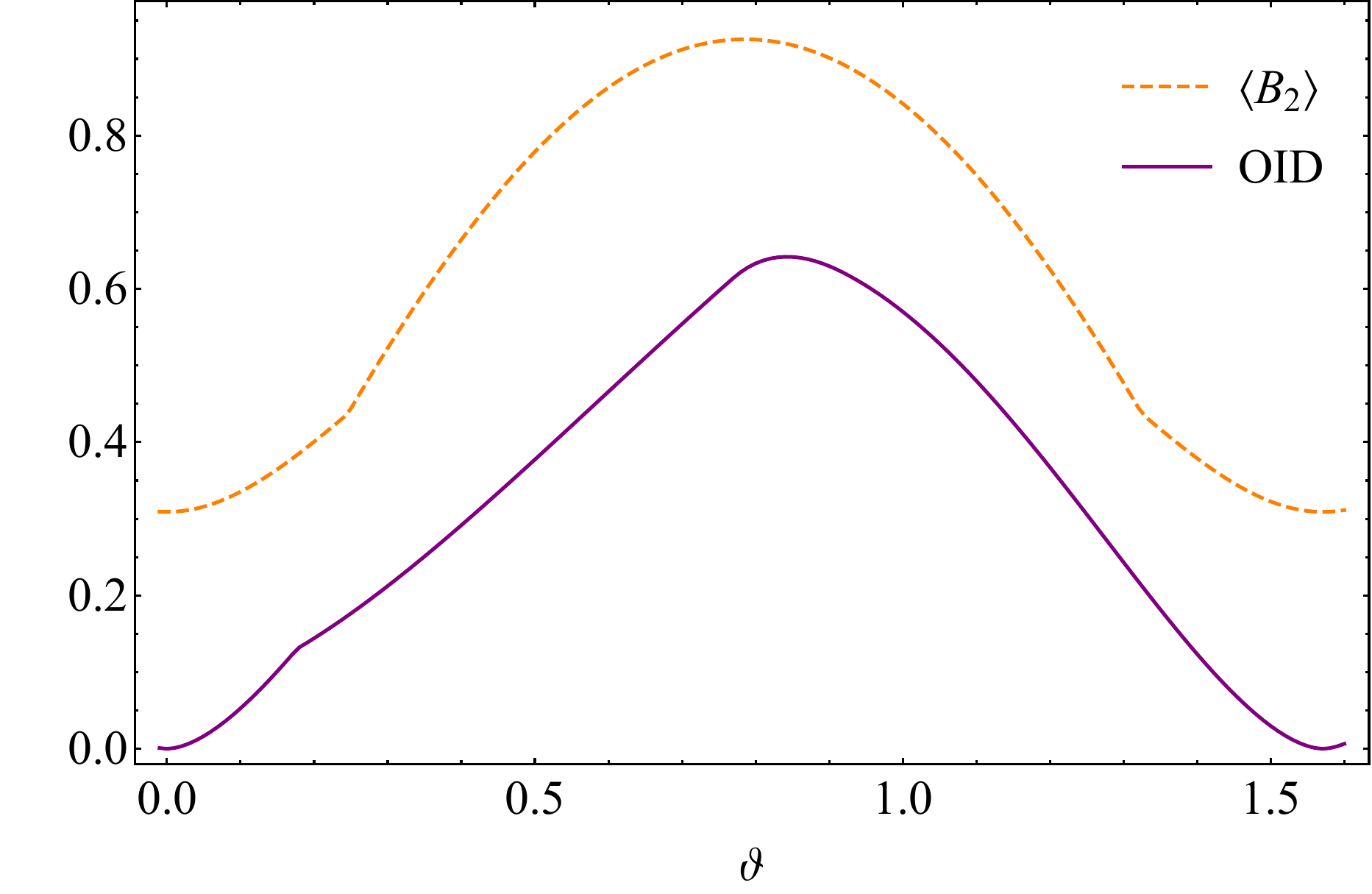}
}
\caption{OID (purple line) and $\langle B_2\rangle$ (dashed orange line) with respect to $\vartheta$ for $\varphi=\pi/5$ for states $\rho_{A,I}$, $\rho_{A,II}$ and $\rho_{I,II}$.
}
\label{bi}
\end{figure}

We plotted the numerical results for both the mean value $\langle B_2\rangle$ and the OID as a function of $\vartheta$. Numerical results for the three
subfigures in Figure \ref{bi} show the $\langle B_2\rangle$ and OID of $\rho_{A,I}$, $\rho_{A,II}$ and $\rho_{I,II}$ as functions of $\vartheta$ for $\varphi=\pi/5$, respectively.
It is seen that that some values of $\langle B_2\rangle$ are
greater than 1. One finds that in Figure \ref{bi} (a) and Figure \ref{bi} (b), there exists
region of $\langle B_2 \rangle$ that violates the Bell inequality. However, in Figure \ref{bi} (c) there is no such violation at all.
In Figure \ref{bi} (a), the OID decreases as $\vartheta$ grows.
In Figure \ref{bi} (b), the ID increases as $\vartheta$ grows.
In Figure \ref{bi} (c), the OID first grows and then decreases.
All $\langle B_2\rangle$ are greater than the OID.
Moreover, the OID doe not vanish in some regions without Bell nonlocality.
This also shows how the acceleration affects the Bell
nonlocality and the one-way quantum deficit.

The main results are presented in Table \ref{tab}. We numerically verify the BMK inequalities and provide analytical or numerical results for the OID of specific states. Similar to quantum discord which has been proven to be NP-complete to compute \cite{Huang2014}, the connection between quantum discord and OID is demonstrated in author's work \cite{Ye2016a}. Consequently, only partial analytical results regarding the interplay between acceleration and quantum correlations for specific cases are presented. It has been shown that the acceleration affects both
Bell nonlocality and the one-way quantum deficit.

\begin{table}[htbp]
    \centering
    \begin{tabular}{|l|l|l|l|}
        \hline
        \textbf{System} & \textbf{Specific state parameter} & \textbf{Optimal measurement parameter} & \textbf{Results} \\
        \hline
        \multirow{3}{*}{$\rho_{A,I}$} & $\varphi=\pi/4$ & $\eta=\pi/2$ & Eq.(\ref{17}) \\ \cline{2-4}
        & $\varphi=\pi/5$ & $\vartheta$-dependent & Numerical results \\ \cline{2-4}
        & $\varphi=\pi/8$ & $\eta=0$ or $\pi$ & Eq.(\ref{18}) \\ \cline{2-4}
        \hline
        \multirow{3}{*}{$\rho_{A,II}$}& $\varphi=\pi/4$ & $\eta=\pi/2$ & Eq.(\ref{20}) \\ \cline{2-4}
        & $\varphi=\pi/5$ & $\vartheta$-dependent & Numerical results \\ \cline{2-4}
        & $\varphi=\pi/8$ & $\eta=0$ or $\pi$ & Eq.(\ref{21}) \\
        \hline
        \multirow{3}{*}{$\rho_{I,II}$} & \multirow{3}{*}{$\varphi=\pi/4, \pi/5, \pi/8$} & \multirow{3}{*}{$\vartheta$-dependent} & \multirow{3}{*}{Numerical results} \\ 
        &  &  &  \\ 
        &  &  &  \\
        \hline
    \end{tabular}
    \caption{One-way Information Deficit (OID) for different systems $\rho_{A,I}$, $\rho_{A,II}$, and $\rho_{I,II}$, with analytical and numerical results for specific
    states.}
    \label{tab}
\end{table}

\section{Conclusions}
We have studied the Bell-Mermin-Klyshko nonlocality and the
one-way information deficit in the Unruh effect in the noninertial system.
Analytical solutions for the OID have been derived for specific states. 
It has been shown that the acceleration has an impact on the Bell
nonlocality and the one-way quantum deficit, which 
reveal the correlations between the inertial and the accelerated
observers in the presence of the Unruh effect. The study provides Bell nonlocality
and one-way information deficit in the context of relativistic quantum information.

\section*{Acknowledgments}
This work was supported in part by
the National Natural Science Foundation of China under Grant Nos. (12075159, 12171044 and 11905131), the Academician Innovation Platform of Hainan Province. This work was also supported by the Jiangxi Provincial Natural Science Foundation of China, under Grants No. 20213BCJL22054.


\begin{thebibliography}{55}%
\makeatletter
\providecommand \@ifxundefined [1]{%
 \@ifx{#1\undefined}
}%
\providecommand \@ifnum [1]{%
 \ifnum #1\expandafter \@firstoftwo
 \else \expandafter \@secondoftwo
 \fi
}%
\providecommand \@ifx [1]{%
 \ifx #1\expandafter \@firstoftwo
 \else \expandafter \@secondoftwo
 \fi
}%
\providecommand \natexlab [1]{#1}%
\providecommand \enquote  [1]{``#1''}%
\providecommand \bibnamefont  [1]{#1}%
\providecommand \bibfnamefont [1]{#1}%
\providecommand \citenamefont [1]{#1}%
\providecommand \href@noop [0]{\@secondoftwo}%
\providecommand \href [0]{\begingroup \@sanitize@url \@href}%
\providecommand \@href[1]{\@@startlink{#1}\@@href}%
\providecommand \@@href[1]{\endgroup#1\@@endlink}%
\providecommand \@sanitize@url [0]{\catcode `\\12\catcode `\$12\catcode
  `\&12\catcode `\#12\catcode `\^12\catcode `\_12\catcode `\%12\relax}%
\providecommand \@@startlink[1]{}%
\providecommand \@@endlink[0]{}%
\providecommand \url  [0]{\begingroup\@sanitize@url \@url }%
\providecommand \@url [1]{\endgroup\@href {#1}{\urlprefix }}%
\providecommand \urlprefix  [0]{URL }%
\providecommand \Eprint [0]{\href }%
\providecommand \doibase [0]{http://dx.doi.org/}%
\providecommand \selectlanguage [0]{\@gobble}%
\providecommand \bibinfo  [0]{\@secondoftwo}%
\providecommand \bibfield  [0]{\@secondoftwo}%
\providecommand \translation [1]{[#1]}%
\providecommand \BibitemOpen [0]{}%
\providecommand \bibitemStop [0]{}%
\providecommand \bibitemNoStop [0]{.\EOS\space}%
\providecommand \EOS [0]{\spacefactor3000\relax}%
\providecommand \BibitemShut  [1]{\csname bibitem#1\endcsname}%
\let\auto@bib@innerbib\@empty
\bibitem [{\citenamefont {Bennett}\ \emph {et~al.}(1993)\citenamefont
  {Bennett}, \citenamefont {Brassard}, \citenamefont {Cr{\'e}peau},
  \citenamefont {Jozsa}, \citenamefont {Peres},\ and\ \citenamefont
  {Wootters}}]{Bennett1993}%
  \BibitemOpen
  \bibfield  {author} {\bibinfo {author} {\bibfnamefont {C.~H.}\ \bibnamefont
  {Bennett}}, \bibinfo {author} {\bibfnamefont {G.}~\bibnamefont {Brassard}},
  \bibinfo {author} {\bibfnamefont {C.}~\bibnamefont {Cr{\'e}peau}}, \bibinfo
  {author} {\bibfnamefont {R.}~\bibnamefont {Jozsa}}, \bibinfo {author}
  {\bibfnamefont {A.}~\bibnamefont {Peres}}, \ and\ \bibinfo {author}
  {\bibfnamefont {W.~K.}\ \bibnamefont {Wootters}},\ }\href@noop {} {\bibfield
  {journal} {\bibinfo  {journal} {Phys. Rev. Lett.}\ }\textbf {\bibinfo
  {volume} {70}},\ \bibinfo {pages} {1895} (\bibinfo {year}
  {1993})}\BibitemShut {NoStop}%
\bibitem [{\citenamefont {Bennett}\ \emph {et~al.}(2001)\citenamefont
  {Bennett}, \citenamefont {DiVincenzo}, \citenamefont {Shor}, \citenamefont
  {Smolin}, \citenamefont {Terhal},\ and\ \citenamefont
  {Wootters}}]{Bennett2001}%
  \BibitemOpen
  \bibfield  {author} {\bibinfo {author} {\bibfnamefont {C.~H.}\ \bibnamefont
  {Bennett}}, \bibinfo {author} {\bibfnamefont {D.~P.}\ \bibnamefont
  {DiVincenzo}}, \bibinfo {author} {\bibfnamefont {P.~W.}\ \bibnamefont
  {Shor}}, \bibinfo {author} {\bibfnamefont {J.~A.}\ \bibnamefont {Smolin}},
  \bibinfo {author} {\bibfnamefont {B.~M.}\ \bibnamefont {Terhal}}, \ and\
  \bibinfo {author} {\bibfnamefont {W.~K.}\ \bibnamefont {Wootters}},\ }\href
  {\doibase 10.1103/PhysRevLett.87.077902} {\bibfield  {journal} {\bibinfo
  {journal} {Phys. Rev. Lett.}\ }\textbf {\bibinfo {volume} {87}},\ \bibinfo
  {pages} {077902} (\bibinfo {year} {2001})}\BibitemShut {NoStop}%
\bibitem [{\citenamefont {Scarani}\ \emph {et~al.}(2009)\citenamefont
  {Scarani}, \citenamefont {Bechmann-Pasquinucci}, \citenamefont {Cerf},
  \citenamefont {Du{\v{s}}ek}, \citenamefont {L{\"u}tkenhaus},\ and\
  \citenamefont {Peev}}]{Scarani2009}%
  \BibitemOpen
  \bibfield  {author} {\bibinfo {author} {\bibfnamefont {V.}~\bibnamefont
  {Scarani}}, \bibinfo {author} {\bibfnamefont {H.}~\bibnamefont
  {Bechmann-Pasquinucci}}, \bibinfo {author} {\bibfnamefont {N.~J.}\
  \bibnamefont {Cerf}}, \bibinfo {author} {\bibfnamefont {M.}~\bibnamefont
  {Du{\v{s}}ek}}, \bibinfo {author} {\bibfnamefont {N.}~\bibnamefont
  {L{\"u}tkenhaus}}, \ and\ \bibinfo {author} {\bibfnamefont {M.}~\bibnamefont
  {Peev}},\ }\href@noop {} {\bibfield  {journal} {\bibinfo  {journal} {Rev.
  Mod. Phys.}\ }\textbf {\bibinfo {volume} {81}},\ \bibinfo {pages} {1301}
  (\bibinfo {year} {2009})}\BibitemShut {NoStop}%
\bibitem [{\citenamefont {Wang}\ \emph {et~al.}(2005)\citenamefont {Wang},
  \citenamefont {Deng}, \citenamefont {Li}, \citenamefont {Liu},\ and\
  \citenamefont {Long}}]{Wang2005}%
  \BibitemOpen
  \bibfield  {author} {\bibinfo {author} {\bibfnamefont {C.}~\bibnamefont
  {Wang}}, \bibinfo {author} {\bibfnamefont {F.-G.}\ \bibnamefont {Deng}},
  \bibinfo {author} {\bibfnamefont {Y.-S.}\ \bibnamefont {Li}}, \bibinfo
  {author} {\bibfnamefont {X.-S.}\ \bibnamefont {Liu}}, \ and\ \bibinfo
  {author} {\bibfnamefont {G.~L.}\ \bibnamefont {Long}},\ }\href {\doibase
  10.1103/PhysRevA.71.044305} {\bibfield  {journal} {\bibinfo  {journal} {Phys.
  Rev. A}\ }\textbf {\bibinfo {volume} {71}},\ \bibinfo {pages} {044305}
  (\bibinfo {year} {2005})}\BibitemShut {NoStop}%
\bibitem [{\citenamefont {Peres}\ and\ \citenamefont
  {Terno}(2004)}]{Peres2004}%
  \BibitemOpen
  \bibfield  {author} {\bibinfo {author} {\bibfnamefont {A.}~\bibnamefont
  {Peres}}\ and\ \bibinfo {author} {\bibfnamefont {D.~R.}\ \bibnamefont
  {Terno}},\ }\href {\doibase 10.1103/RevModPhys.76.93} {\bibfield  {journal}
  {\bibinfo  {journal} {Rev. Mod. Phys.}\ }\textbf {\bibinfo {volume} {76}},\
  \bibinfo {pages} {93} (\bibinfo {year} {2004})}\BibitemShut {NoStop}%
\bibitem [{\citenamefont {Fuentes-Schuller}\ and\ \citenamefont
  {Mann}(2005)}]{Fuentes-Schuller2005}%
  \BibitemOpen
  \bibfield  {author} {\bibinfo {author} {\bibfnamefont {I.}~\bibnamefont
  {Fuentes-Schuller}}\ and\ \bibinfo {author} {\bibfnamefont {R.~B.}\
  \bibnamefont {Mann}},\ }\href {\doibase 10.1103/PhysRevLett.95.120404}
  {\bibfield  {journal} {\bibinfo  {journal} {Phys. Rev. Lett.}\ }\textbf
  {\bibinfo {volume} {95}},\ \bibinfo {pages} {120404} (\bibinfo {year}
  {2005})}\BibitemShut {NoStop}%
\bibitem [{\citenamefont {Mart{\'{i}}n-Mart{\'{i}}nez}\ and\ \citenamefont
  {Le{\'{o}}n}(2009)}]{Martin-Martinez2009}%
  \BibitemOpen
  \bibfield  {author} {\bibinfo {author} {\bibfnamefont {E.}~\bibnamefont
  {Mart{\'{i}}n-Mart{\'{i}}nez}}\ and\ \bibinfo {author} {\bibfnamefont
  {J.}~\bibnamefont {Le{\'{o}}n}},\ }\href {\doibase
  10.1103/physreva.80.042318} {\bibfield  {journal} {\bibinfo  {journal} {Phys.
  Rev. A}\ }\textbf {\bibinfo {volume} {80}},\ \bibinfo {pages} {042318}
  (\bibinfo {year} {2009})}\BibitemShut {NoStop}%
\bibitem [{\citenamefont {Mart{\'{i}}n-Mart{\'{i}}nez}\ \emph
  {et~al.}(2010)\citenamefont {Mart{\'{i}}n-Mart{\'{i}}nez}, \citenamefont
  {Garay},\ and\ \citenamefont {Le{\'{o}}n}}]{Martin-Martinez2010b}%
  \BibitemOpen
  \bibfield  {author} {\bibinfo {author} {\bibfnamefont {E.}~\bibnamefont
  {Mart{\'{i}}n-Mart{\'{i}}nez}}, \bibinfo {author} {\bibfnamefont {L.~J.}\
  \bibnamefont {Garay}}, \ and\ \bibinfo {author} {\bibfnamefont
  {J.}~\bibnamefont {Le{\'{o}}n}},\ }\href {\doibase
  10.1103/PhysRevD.82.064006} {\bibfield  {journal} {\bibinfo  {journal} {Phys.
  Rev. D}\ }\textbf {\bibinfo {volume} {82}},\ \bibinfo {pages} {064006}
  (\bibinfo {year} {2010})}\BibitemShut {NoStop}%
\bibitem [{\citenamefont {Alsing}\ \emph {et~al.}(2006)\citenamefont {Alsing},
  \citenamefont {Fuentes-Schuller}, \citenamefont {Mann},\ and\ \citenamefont
  {Tessier}}]{Alsing2006}%
  \BibitemOpen
  \bibfield  {author} {\bibinfo {author} {\bibfnamefont {P.~M.}\ \bibnamefont
  {Alsing}}, \bibinfo {author} {\bibfnamefont {I.}~\bibnamefont
  {Fuentes-Schuller}}, \bibinfo {author} {\bibfnamefont {R.~B.}\ \bibnamefont
  {Mann}}, \ and\ \bibinfo {author} {\bibfnamefont {T.~E.}\ \bibnamefont
  {Tessier}},\ }\href {\doibase 10.1103/PhysRevA.74.032326} {\bibfield
  {journal} {\bibinfo  {journal} {Phys. Rev. A}\ }\textbf {\bibinfo {volume}
  {74}},\ \bibinfo {pages} {032326} (\bibinfo {year} {2006})}\BibitemShut
  {NoStop}%
\bibitem [{\citenamefont {Mart{\'{i}}n-Mart{\'{i}}nez}\ and\ \citenamefont
  {Fuentes}(2011)}]{Martin-Martinez2011}%
  \BibitemOpen
  \bibfield  {author} {\bibinfo {author} {\bibfnamefont {E.}~\bibnamefont
  {Mart{\'{i}}n-Mart{\'{i}}nez}}\ and\ \bibinfo {author} {\bibfnamefont
  {I.}~\bibnamefont {Fuentes}},\ }\href {\doibase 10.1103/PhysRevA.83.052306}
  {\bibfield  {journal} {\bibinfo  {journal} {Phys. Rev. A}\ }\textbf {\bibinfo
  {volume} {83}},\ \bibinfo {pages} {052306} (\bibinfo {year}
  {2011})}\BibitemShut {NoStop}%
\bibitem [{\citenamefont {Hwang}\ \emph {et~al.}(2011)\citenamefont {Hwang},
  \citenamefont {Park},\ and\ \citenamefont {Jung}}]{Hwang2011}%
  \BibitemOpen
  \bibfield  {author} {\bibinfo {author} {\bibfnamefont {M.~R.}\ \bibnamefont
  {Hwang}}, \bibinfo {author} {\bibfnamefont {D.}~\bibnamefont {Park}}, \ and\
  \bibinfo {author} {\bibfnamefont {E.}~\bibnamefont {Jung}},\ }\href {\doibase
  10.1103/PhysRevA.83.012111} {\bibfield  {journal} {\bibinfo  {journal} {Phys.
  Rev. A}\ }\textbf {\bibinfo {volume} {83}},\ \bibinfo {pages} {012111}
  (\bibinfo {year} {2011})}\BibitemShut {NoStop}%
\bibitem [{\citenamefont {Friis}\ \emph {et~al.}(2011)\citenamefont {Friis},
  \citenamefont {K{\"{o}}hler}, \citenamefont {Mart{\'{i}}n-Mart{\'{i}}nez},\
  and\ \citenamefont {Bertlmann}}]{Friis2011}%
  \BibitemOpen
  \bibfield  {author} {\bibinfo {author} {\bibfnamefont {N.}~\bibnamefont
  {Friis}}, \bibinfo {author} {\bibfnamefont {P.}~\bibnamefont {K{\"{o}}hler}},
  \bibinfo {author} {\bibfnamefont {E.}~\bibnamefont
  {Mart{\'{i}}n-Mart{\'{i}}nez}}, \ and\ \bibinfo {author} {\bibfnamefont
  {R.~A.}\ \bibnamefont {Bertlmann}},\ }\href {\doibase
  10.1103/PhysRevA.84.062111} {\bibfield  {journal} {\bibinfo  {journal} {Phys.
  Rev. A}\ }\textbf {\bibinfo {volume} {84}},\ \bibinfo {pages} {062111}
  (\bibinfo {year} {2011})}\BibitemShut {NoStop}%
\bibitem [{\citenamefont {Montero}\ \emph {et~al.}(2011)\citenamefont
  {Montero}, \citenamefont {Le{\'{o}}n},\ and\ \citenamefont
  {Mart{\'{i}}n-Mart{\'{i}}nez}}]{Montero2011}%
  \BibitemOpen
  \bibfield  {author} {\bibinfo {author} {\bibfnamefont {M.}~\bibnamefont
  {Montero}}, \bibinfo {author} {\bibfnamefont {J.}~\bibnamefont {Le{\'{o}}n}},
  \ and\ \bibinfo {author} {\bibfnamefont {E.}~\bibnamefont
  {Mart{\'{i}}n-Mart{\'{i}}nez}},\ }\href {\doibase 10.1103/PhysRevA.84.042320}
  {\bibfield  {journal} {\bibinfo  {journal} {Phys. Rev. A}\ }\textbf {\bibinfo
  {volume} {84}},\ \bibinfo {pages} {042320} (\bibinfo {year}
  {2011})}\BibitemShut {NoStop}%
\bibitem [{\citenamefont {Chang}\ and\ \citenamefont {Kwon}(2012)}]{Chang2012}%
  \BibitemOpen
  \bibfield  {author} {\bibinfo {author} {\bibfnamefont {J.}~\bibnamefont
  {Chang}}\ and\ \bibinfo {author} {\bibfnamefont {Y.}~\bibnamefont {Kwon}},\
  }\href {\doibase 10.1103/PhysRevA.85.032302} {\bibfield  {journal} {\bibinfo
  {journal} {Phys. Rev. A}\ }\textbf {\bibinfo {volume} {85}},\ \bibinfo
  {pages} {032302} (\bibinfo {year} {2012})}\BibitemShut {NoStop}%
\bibitem [{\citenamefont {Montero}\ and\ \citenamefont
  {Mart{\'{i}}n-Mart{\'{i}}nez}(2012)}]{Montero2012}%
  \BibitemOpen
  \bibfield  {author} {\bibinfo {author} {\bibfnamefont {M.}~\bibnamefont
  {Montero}}\ and\ \bibinfo {author} {\bibfnamefont {E.}~\bibnamefont
  {Mart{\'{i}}n-Mart{\'{i}}nez}},\ }\href {\doibase 10.1103/PhysRevA.85.024301}
  {\bibfield  {journal} {\bibinfo  {journal} {Phys. Rev. A}\ }\textbf {\bibinfo
  {volume} {85}},\ \bibinfo {pages} {024301} (\bibinfo {year}
  {2012})}\BibitemShut {NoStop}%
\bibitem [{\citenamefont {Wang}\ \emph {et~al.}(2020)\citenamefont {Wang},
  \citenamefont {Liang},\ and\ \citenamefont {Zheng}}]{Wang2020}%
  \BibitemOpen
  \bibfield  {author} {\bibinfo {author} {\bibfnamefont {K.}~\bibnamefont
  {Wang}}, \bibinfo {author} {\bibfnamefont {Y.}~\bibnamefont {Liang}}, \ and\
  \bibinfo {author} {\bibfnamefont {Z.~J.}\ \bibnamefont {Zheng}},\ }\href
  {\doibase 10.1007/s11128-020-02645-1} {\bibfield  {journal} {\bibinfo
  {journal} {Quantum Inf. Process.}\ }\textbf {\bibinfo {volume} {19}},\
  \bibinfo {pages} {140} (\bibinfo {year} {2020})}\BibitemShut {NoStop}%
\bibitem [{\citenamefont {Alsing}\ and\ \citenamefont
  {Milburn}(2003)}]{Alsing2003}%
  \BibitemOpen
  \bibfield  {author} {\bibinfo {author} {\bibfnamefont {P.~M.}\ \bibnamefont
  {Alsing}}\ and\ \bibinfo {author} {\bibfnamefont {G.~J.}\ \bibnamefont
  {Milburn}},\ }\href {\doibase 10.1103/PhysRevLett.91.180404} {\bibfield
  {journal} {\bibinfo  {journal} {Phys. Rev. Lett.}\ }\textbf {\bibinfo
  {volume} {91}},\ \bibinfo {pages} {180404} (\bibinfo {year}
  {2003})}\BibitemShut {NoStop}%
\bibitem [{\citenamefont {Mart{\'{i}}n-Mart{\'{i}}nez}\ and\ \citenamefont
  {Le{\'{o}}n}(2010{\natexlab{a}})}]{Martin-Martinez2010}%
  \BibitemOpen
  \bibfield  {author} {\bibinfo {author} {\bibfnamefont {E.}~\bibnamefont
  {Mart{\'{i}}n-Mart{\'{i}}nez}}\ and\ \bibinfo {author} {\bibfnamefont
  {J.}~\bibnamefont {Le{\'{o}}n}},\ }\href {\doibase
  10.1103/PhysRevA.81.052305} {\bibfield  {journal} {\bibinfo  {journal} {Phys.
  Rev. A}\ }\textbf {\bibinfo {volume} {81}},\ \bibinfo {pages} {052305}
  (\bibinfo {year} {2010}{\natexlab{a}})}\BibitemShut {NoStop}%
\bibitem [{\citenamefont {Mart{\'{i}}n-Mart{\'{i}}nez}\ and\ \citenamefont
  {Le{\'{o}}n}(2010{\natexlab{b}})}]{Martin-Martinez2010a}%
  \BibitemOpen
  \bibfield  {author} {\bibinfo {author} {\bibfnamefont {E.}~\bibnamefont
  {Mart{\'{i}}n-Mart{\'{i}}nez}}\ and\ \bibinfo {author} {\bibfnamefont
  {J.}~\bibnamefont {Le{\'{o}}n}},\ }\href {\doibase
  10.1103/PhysRevA.81.032320} {\bibfield  {journal} {\bibinfo  {journal} {Phys.
  Rev. A}\ }\textbf {\bibinfo {volume} {81}},\ \bibinfo {pages} {032320}
  (\bibinfo {year} {2010}{\natexlab{b}})}\BibitemShut {NoStop}%
\bibitem [{\citenamefont {Bruschi}\ \emph {et~al.}(2010)\citenamefont
  {Bruschi}, \citenamefont {Louko}, \citenamefont
  {Mart{\'{i}}n-Mart{\'{i}}nez}, \citenamefont {Dragan},\ and\ \citenamefont
  {Fuentes}}]{Bruschi2010}%
  \BibitemOpen
  \bibfield  {author} {\bibinfo {author} {\bibfnamefont {D.~E.}\ \bibnamefont
  {Bruschi}}, \bibinfo {author} {\bibfnamefont {J.}~\bibnamefont {Louko}},
  \bibinfo {author} {\bibfnamefont {E.}~\bibnamefont
  {Mart{\'{i}}n-Mart{\'{i}}nez}}, \bibinfo {author} {\bibfnamefont
  {A.}~\bibnamefont {Dragan}}, \ and\ \bibinfo {author} {\bibfnamefont
  {I.}~\bibnamefont {Fuentes}},\ }\href {\doibase 10.1103/PhysRevA.82.042332}
  {\bibfield  {journal} {\bibinfo  {journal} {Phys. Rev. A}\ }\textbf {\bibinfo
  {volume} {82}},\ \bibinfo {pages} {042332} (\bibinfo {year}
  {2010})}\BibitemShut {NoStop}%
\bibitem [{\citenamefont {Montero}\ and\ \citenamefont
  {Mart{\'{i}}n-Mart{\'{i}}nez}(2011)}]{Montero2011a}%
  \BibitemOpen
  \bibfield  {author} {\bibinfo {author} {\bibfnamefont {M.}~\bibnamefont
  {Montero}}\ and\ \bibinfo {author} {\bibfnamefont {E.}~\bibnamefont
  {Mart{\'{i}}n-Mart{\'{i}}nez}},\ }\href@noop {} {\bibfield  {journal}
  {\bibinfo  {journal} {J. High Energy Phys.}\ }\textbf {\bibinfo {volume}
  {07}},\ \bibinfo {pages} {006} (\bibinfo {year} {2011})}\BibitemShut
  {NoStop}%
\bibitem [{\citenamefont {Ahmadi}\ \emph {et~al.}(2016)\citenamefont {Ahmadi},
  \citenamefont {Lorek}, \citenamefont {Ch{\k{e}}ci{\'n}ska}, \citenamefont
  {Smith}, \citenamefont {Mann},\ and\ \citenamefont {Dragan}}]{Ahmadi2016}%
  \BibitemOpen
  \bibfield  {author} {\bibinfo {author} {\bibfnamefont {M.}~\bibnamefont
  {Ahmadi}}, \bibinfo {author} {\bibfnamefont {K.}~\bibnamefont {Lorek}},
  \bibinfo {author} {\bibfnamefont {A.}~\bibnamefont {Ch{\k{e}}ci{\'n}ska}},
  \bibinfo {author} {\bibfnamefont {A.~R.}\ \bibnamefont {Smith}}, \bibinfo
  {author} {\bibfnamefont {R.~B.}\ \bibnamefont {Mann}}, \ and\ \bibinfo
  {author} {\bibfnamefont {A.}~\bibnamefont {Dragan}},\ }\href@noop {}
  {\bibfield  {journal} {\bibinfo  {journal} {Phys. Rev. D}\ }\textbf {\bibinfo
  {volume} {93}},\ \bibinfo {pages} {124031} (\bibinfo {year}
  {2016})}\BibitemShut {NoStop}%
\bibitem [{\citenamefont {Xu}\ \emph {et~al.}(2020)\citenamefont {Xu},
  \citenamefont {Zhu}, \citenamefont {Zhang}, \citenamefont {Wang},\ and\
  \citenamefont {Liu}}]{Xu2020}%
  \BibitemOpen
  \bibfield  {author} {\bibinfo {author} {\bibfnamefont {K.}~\bibnamefont
  {Xu}}, \bibinfo {author} {\bibfnamefont {H.~J.}\ \bibnamefont {Zhu}},
  \bibinfo {author} {\bibfnamefont {G.~F.}\ \bibnamefont {Zhang}}, \bibinfo
  {author} {\bibfnamefont {J.~C.}\ \bibnamefont {Wang}}, \ and\ \bibinfo
  {author} {\bibfnamefont {W.~M.}\ \bibnamefont {Liu}},\ }\href {\doibase
  10.1140/epjc/s10052-020-8048-x} {\bibfield  {journal} {\bibinfo  {journal}
  {Eur. Phys. J. C}\ }\textbf {\bibinfo {volume} {80}},\ \bibinfo {pages} {462}
  (\bibinfo {year} {2020})}\BibitemShut {NoStop}%
\bibitem [{\citenamefont {Li}\ \emph {et~al.}(2022)\citenamefont {Li},
  \citenamefont {Ming}, \citenamefont {Song}, \citenamefont {Ye},\ and\
  \citenamefont {Wang}}]{Li2022}%
  \BibitemOpen
  \bibfield  {author} {\bibinfo {author} {\bibfnamefont {L.~J.}\ \bibnamefont
  {Li}}, \bibinfo {author} {\bibfnamefont {F.}~\bibnamefont {Ming}}, \bibinfo
  {author} {\bibfnamefont {X.~K.}\ \bibnamefont {Song}}, \bibinfo {author}
  {\bibfnamefont {L.}~\bibnamefont {Ye}}, \ and\ \bibinfo {author}
  {\bibfnamefont {D.}~\bibnamefont {Wang}},\ }\href {\doibase
  10.1140/epjc/s10052-022-10687-1} {\bibfield  {journal} {\bibinfo  {journal}
  {Eur. Phys. J. C}\ }\textbf {\bibinfo {volume} {82}},\ \bibinfo {pages} {726}
  (\bibinfo {year} {2022})}\BibitemShut {NoStop}%
\bibitem [{\citenamefont {Liu}\ \emph {et~al.}(2023)\citenamefont {Liu},
  \citenamefont {Wu}, \citenamefont {Wen},\ and\ \citenamefont
  {Wang}}]{Liu2023}%
  \BibitemOpen
  \bibfield  {author} {\bibinfo {author} {\bibfnamefont {Q.}~\bibnamefont
  {Liu}}, \bibinfo {author} {\bibfnamefont {S.~M.}\ \bibnamefont {Wu}},
  \bibinfo {author} {\bibfnamefont {C.}~\bibnamefont {Wen}}, \ and\ \bibinfo
  {author} {\bibfnamefont {J.}~\bibnamefont {Wang}},\ }\href {\doibase
  10.1007/s11433-023-2246-8} {\bibfield  {journal} {\bibinfo  {journal} {Sci.
  China-Phys. Mech. Astron.}\ }\textbf {\bibinfo {volume} {66}},\ \bibinfo
  {pages} {120413} (\bibinfo {year} {2023})}\BibitemShut {NoStop}%
\bibitem [{\citenamefont {Zhang}\ \emph {et~al.}(2023)\citenamefont {Zhang},
  \citenamefont {Wang},\ and\ \citenamefont {Fei}}]{Zhang2023}%
  \BibitemOpen
  \bibfield  {author} {\bibinfo {author} {\bibfnamefont {T.}~\bibnamefont
  {Zhang}}, \bibinfo {author} {\bibfnamefont {X.}~\bibnamefont {Wang}}, \ and\
  \bibinfo {author} {\bibfnamefont {S.-M.}\ \bibnamefont {Fei}},\ }\href
  {\doibase 10.1140/epjc/s10052-023-11796-1} {\bibfield  {journal} {\bibinfo
  {journal} {Eur. Phys. J. C}\ }\textbf {\bibinfo {volume} {83}},\ \bibinfo
  {pages} {607} (\bibinfo {year} {2023})}\BibitemShut {NoStop}%
\bibitem [{\citenamefont {Wang}\ and\ \citenamefont {Jing}(2010)}]{Wang2010a}%
  \BibitemOpen
  \bibfield  {author} {\bibinfo {author} {\bibfnamefont {J.}~\bibnamefont
  {Wang}}\ and\ \bibinfo {author} {\bibfnamefont {J.}~\bibnamefont {Jing}},\
  }\href {\doibase 10.1103/PhysRevA.82.032324} {\bibfield  {journal} {\bibinfo
  {journal} {Phys. Rev. A}\ }\textbf {\bibinfo {volume} {82}},\ \bibinfo
  {pages} {032324} (\bibinfo {year} {2010})}\BibitemShut {NoStop}%
\bibitem [{\citenamefont {Xiao}\ \emph {et~al.}(2018)\citenamefont {Xiao},
  \citenamefont {Xie}, \citenamefont {Yao}, \citenamefont {Li},\ and\
  \citenamefont {Wang}}]{Xiao2018}%
  \BibitemOpen
  \bibfield  {author} {\bibinfo {author} {\bibfnamefont {X.}~\bibnamefont
  {Xiao}}, \bibinfo {author} {\bibfnamefont {Y.-M.}\ \bibnamefont {Xie}},
  \bibinfo {author} {\bibfnamefont {Y.}~\bibnamefont {Yao}}, \bibinfo {author}
  {\bibfnamefont {Y.-L.}\ \bibnamefont {Li}}, \ and\ \bibinfo {author}
  {\bibfnamefont {J.}~\bibnamefont {Wang}},\ }\href@noop {} {\bibfield
  {journal} {\bibinfo  {journal} {Ann. Phys.}\ }\textbf {\bibinfo {volume}
  {390}},\ \bibinfo {pages} {83} (\bibinfo {year} {2018})}\BibitemShut
  {NoStop}%
\bibitem [{\citenamefont {Ollivier}\ and\ \citenamefont
  {Zurek}(2001)}]{Ollivier2001}%
  \BibitemOpen
  \bibfield  {author} {\bibinfo {author} {\bibfnamefont {H.}~\bibnamefont
  {Ollivier}}\ and\ \bibinfo {author} {\bibfnamefont {W.~H.}\ \bibnamefont
  {Zurek}},\ }\href@noop {} {\bibfield  {journal} {\bibinfo  {journal} {Phys.
  Rev. Lett.}\ }\textbf {\bibinfo {volume} {88}},\ \bibinfo {pages} {017901}
  (\bibinfo {year} {2001})}\BibitemShut {NoStop}%
\bibitem [{\citenamefont {Luo}(2008)}]{Luo2008}%
  \BibitemOpen
  \bibfield  {author} {\bibinfo {author} {\bibfnamefont {S.}~\bibnamefont
  {Luo}},\ }\href@noop {} {\bibfield  {journal} {\bibinfo  {journal} {Phys.
  Rev. A}\ }\textbf {\bibinfo {volume} {77}},\ \bibinfo {pages} {022301}
  (\bibinfo {year} {2008})}\BibitemShut {NoStop}%
\bibitem [{\citenamefont {Daki{\'c}}\ \emph {et~al.}(2010)\citenamefont
  {Daki{\'c}}, \citenamefont {Vedral},\ and\ \citenamefont
  {Brukner}}]{Dakic2010}%
  \BibitemOpen
  \bibfield  {author} {\bibinfo {author} {\bibfnamefont {B.}~\bibnamefont
  {Daki{\'c}}}, \bibinfo {author} {\bibfnamefont {V.}~\bibnamefont {Vedral}}, \
  and\ \bibinfo {author} {\bibfnamefont {{\v{C}}.}~\bibnamefont {Brukner}},\
  }\href@noop {} {\bibfield  {journal} {\bibinfo  {journal} {Phys. Rev. Lett.}\
  }\textbf {\bibinfo {volume} {105}},\ \bibinfo {pages} {190502} (\bibinfo
  {year} {2010})}\BibitemShut {NoStop}%
\bibitem [{\citenamefont {Horodecki}\ \emph {et~al.}(2005)\citenamefont
  {Horodecki}, \citenamefont {Horodecki}, \citenamefont {Horodecki},
  \citenamefont {Oppenheim}, \citenamefont {Sen(De)}, \citenamefont {Sen},\
  and\ \citenamefont {Synak-Radtke}}]{Horodecki2005}%
  \BibitemOpen
  \bibfield  {author} {\bibinfo {author} {\bibfnamefont {M.}~\bibnamefont
  {Horodecki}}, \bibinfo {author} {\bibfnamefont {P.}~\bibnamefont
  {Horodecki}}, \bibinfo {author} {\bibfnamefont {R.}~\bibnamefont
  {Horodecki}}, \bibinfo {author} {\bibfnamefont {J.}~\bibnamefont
  {Oppenheim}}, \bibinfo {author} {\bibfnamefont {A.}~\bibnamefont {Sen(De)}},
  \bibinfo {author} {\bibfnamefont {U.}~\bibnamefont {Sen}}, \ and\ \bibinfo
  {author} {\bibfnamefont {B.}~\bibnamefont {Synak-Radtke}},\ }\href {\doibase
  10.1103/PhysRevA.71.062307} {\bibfield  {journal} {\bibinfo  {journal} {Phys.
  Rev. A}\ }\textbf {\bibinfo {volume} {71}},\ \bibinfo {pages} {062307}
  (\bibinfo {year} {2005})}\BibitemShut {NoStop}%
\bibitem [{\citenamefont {Streltsov}\ \emph {et~al.}(2011)\citenamefont
  {Streltsov}, \citenamefont {Kampermann},\ and\ \citenamefont
  {Bru{\ss}}}]{Streltsov2011}%
  \BibitemOpen
  \bibfield  {author} {\bibinfo {author} {\bibfnamefont {A.}~\bibnamefont
  {Streltsov}}, \bibinfo {author} {\bibfnamefont {H.}~\bibnamefont
  {Kampermann}}, \ and\ \bibinfo {author} {\bibfnamefont {D.}~\bibnamefont
  {Bru{\ss}}},\ }\href {\doibase 10.1103/PhysRevLett.106.160401} {\bibfield
  {journal} {\bibinfo  {journal} {Phys. Rev. Lett.}\ }\textbf {\bibinfo
  {volume} {106}},\ \bibinfo {pages} {160401} (\bibinfo {year}
  {2011})}\BibitemShut {NoStop}%
\bibitem [{\citenamefont {Ciliberti}\ \emph {et~al.}(2013)\citenamefont
  {Ciliberti}, \citenamefont {Canosa},\ and\ \citenamefont
  {Rossignoli}}]{Ciliberti2013}%
  \BibitemOpen
  \bibfield  {author} {\bibinfo {author} {\bibfnamefont {L.}~\bibnamefont
  {Ciliberti}}, \bibinfo {author} {\bibfnamefont {N.}~\bibnamefont {Canosa}}, \
  and\ \bibinfo {author} {\bibfnamefont {R.}~\bibnamefont {Rossignoli}},\
  }\href {\doibase 10.1103/PhysRevA.88.012119} {\bibfield  {journal} {\bibinfo
  {journal} {Phys. Rev. A}\ }\textbf {\bibinfo {volume} {88}},\ \bibinfo
  {pages} {012119} (\bibinfo {year} {2013})}\BibitemShut {NoStop}%
\bibitem [{\citenamefont {Wang}\ \emph {et~al.}(2013)\citenamefont {Wang},
  \citenamefont {Ma}, \citenamefont {Li},\ and\ \citenamefont
  {Wang}}]{Wang2013}%
  \BibitemOpen
  \bibfield  {author} {\bibinfo {author} {\bibfnamefont {Y.-K.}\ \bibnamefont
  {Wang}}, \bibinfo {author} {\bibfnamefont {T.}~\bibnamefont {Ma}}, \bibinfo
  {author} {\bibfnamefont {B.}~\bibnamefont {Li}}, \ and\ \bibinfo {author}
  {\bibfnamefont {Z.-X.}\ \bibnamefont {Wang}},\ }\href@noop {} {\bibfield
  {journal} {\bibinfo  {journal} {Commun. Theor. Phys.}\ }\textbf {\bibinfo
  {volume} {59}},\ \bibinfo {pages} {540} (\bibinfo {year} {2013})}\BibitemShut
  {NoStop}%
\bibitem [{\citenamefont {Wang}\ \emph {et~al.}(2015)\citenamefont {Wang},
  \citenamefont {Jing}, \citenamefont {Fei}, \citenamefont {Wang},
  \citenamefont {Cao},\ and\ \citenamefont {Fan}}]{Wang2015}%
  \BibitemOpen
  \bibfield  {author} {\bibinfo {author} {\bibfnamefont {Y.-K.}\ \bibnamefont
  {Wang}}, \bibinfo {author} {\bibfnamefont {N.}~\bibnamefont {Jing}}, \bibinfo
  {author} {\bibfnamefont {S.-M.}\ \bibnamefont {Fei}}, \bibinfo {author}
  {\bibfnamefont {Z.-X.}\ \bibnamefont {Wang}}, \bibinfo {author}
  {\bibfnamefont {J.-P.}\ \bibnamefont {Cao}}, \ and\ \bibinfo {author}
  {\bibfnamefont {H.}~\bibnamefont {Fan}},\ }\href@noop {} {\bibfield
  {journal} {\bibinfo  {journal} {Quantum Inf. Process.}\ }\textbf {\bibinfo
  {volume} {14}},\ \bibinfo {pages} {2487} (\bibinfo {year}
  {2015})}\BibitemShut {NoStop}%
\bibitem [{\citenamefont {Ye}\ \emph {et~al.}(2017)\citenamefont {Ye},
  \citenamefont {Li}, \citenamefont {Zhao}, \citenamefont {Zhang},\ and\
  \citenamefont {Fei}}]{Ye2017}%
  \BibitemOpen
  \bibfield  {author} {\bibinfo {author} {\bibfnamefont {B.-L.}\ \bibnamefont
  {Ye}}, \bibinfo {author} {\bibfnamefont {B.}~\bibnamefont {Li}}, \bibinfo
  {author} {\bibfnamefont {L.-J.}\ \bibnamefont {Zhao}}, \bibinfo {author}
  {\bibfnamefont {H.-J.}\ \bibnamefont {Zhang}}, \ and\ \bibinfo {author}
  {\bibfnamefont {S.-M.}\ \bibnamefont {Fei}},\ }\href@noop {} {\bibfield
  {journal} {\bibinfo  {journal} {Sci. China-Phys. Mech. Astron.}\ }\textbf
  {\bibinfo {volume} {60}},\ \bibinfo {pages} {030311} (\bibinfo {year}
  {2017})}\BibitemShut {NoStop}%
\bibitem [{\citenamefont {Modi}\ \emph {et~al.}(2012)\citenamefont {Modi},
  \citenamefont {Brodutch}, \citenamefont {Cable}, \citenamefont {Paterek},\
  and\ \citenamefont {Vedral}}]{Modi2012}%
  \BibitemOpen
  \bibfield  {author} {\bibinfo {author} {\bibfnamefont {K.}~\bibnamefont
  {Modi}}, \bibinfo {author} {\bibfnamefont {A.}~\bibnamefont {Brodutch}},
  \bibinfo {author} {\bibfnamefont {H.}~\bibnamefont {Cable}}, \bibinfo
  {author} {\bibfnamefont {T.}~\bibnamefont {Paterek}}, \ and\ \bibinfo
  {author} {\bibfnamefont {V.}~\bibnamefont {Vedral}},\ }\href@noop {}
  {\bibfield  {journal} {\bibinfo  {journal} {Rev. Mod. Phys.}\ }\textbf
  {\bibinfo {volume} {84}},\ \bibinfo {pages} {1655} (\bibinfo {year}
  {2012})}\BibitemShut {NoStop}%
\bibitem [{\citenamefont {Adesso}\ \emph {et~al.}(2016)\citenamefont {Adesso},
  \citenamefont {Bromley},\ and\ \citenamefont {Cianciaruso}}]{Adesso2016}%
  \BibitemOpen
  \bibfield  {author} {\bibinfo {author} {\bibfnamefont {G.}~\bibnamefont
  {Adesso}}, \bibinfo {author} {\bibfnamefont {T.~R.}\ \bibnamefont {Bromley}},
  \ and\ \bibinfo {author} {\bibfnamefont {M.}~\bibnamefont {Cianciaruso}},\
  }\href@noop {} {\bibfield  {journal} {\bibinfo  {journal} {J. Phys. A: Math.
  Theor.}\ }\textbf {\bibinfo {volume} {49}},\ \bibinfo {pages} {473001}
  (\bibinfo {year} {2016})}\BibitemShut {NoStop}%
\bibitem [{\citenamefont {Datta}(2009)}]{Datta2009}%
  \BibitemOpen
  \bibfield  {author} {\bibinfo {author} {\bibfnamefont {A.}~\bibnamefont
  {Datta}},\ }\href {\doibase 10.1103/PhysRevA.80.052304} {\bibfield  {journal}
  {\bibinfo  {journal} {Phys. Rev. A}\ }\textbf {\bibinfo {volume} {80}},\
  \bibinfo {pages} {052304} (\bibinfo {year} {2009})}\BibitemShut {NoStop}%
\bibitem [{\citenamefont {Wang}\ \emph {et~al.}(2010)\citenamefont {Wang},
  \citenamefont {Deng},\ and\ \citenamefont {Jing}}]{Wang2010}%
  \BibitemOpen
  \bibfield  {author} {\bibinfo {author} {\bibfnamefont {J.}~\bibnamefont
  {Wang}}, \bibinfo {author} {\bibfnamefont {J.}~\bibnamefont {Deng}}, \ and\
  \bibinfo {author} {\bibfnamefont {J.}~\bibnamefont {Jing}},\ }\href {\doibase
  10.1103/PhysRevA.81.052120} {\bibfield  {journal} {\bibinfo  {journal} {Phys.
  Rev. A}\ }\textbf {\bibinfo {volume} {81}},\ \bibinfo {pages} {052120}
  (\bibinfo {year} {2010})}\BibitemShut {NoStop}%
\bibitem [{\citenamefont {Mehri-Dehnavi}\ \emph {et~al.}(2011)\citenamefont
  {Mehri-Dehnavi}, \citenamefont {Mirza}, \citenamefont {Mohammadzadeh},\ and\
  \citenamefont {Rahimi}}]{Mehri-Dehnavi2011}%
  \BibitemOpen
  \bibfield  {author} {\bibinfo {author} {\bibfnamefont {H.}~\bibnamefont
  {Mehri-Dehnavi}}, \bibinfo {author} {\bibfnamefont {B.}~\bibnamefont
  {Mirza}}, \bibinfo {author} {\bibfnamefont {H.}~\bibnamefont
  {Mohammadzadeh}}, \ and\ \bibinfo {author} {\bibfnamefont {R.}~\bibnamefont
  {Rahimi}},\ }\href {\doibase 10.1016/j.aop.2011.02.001} {\bibfield  {journal}
  {\bibinfo  {journal} {Ann. Phys.}\ }\textbf {\bibinfo {volume} {326}},\
  \bibinfo {pages} {1320} (\bibinfo {year} {2011})}\BibitemShut {NoStop}%
\bibitem [{\citenamefont {Wang}\ \emph {et~al.}(2014)\citenamefont {Wang},
  \citenamefont {Jing},\ and\ \citenamefont {Fan}}]{Wang2014}%
  \BibitemOpen
  \bibfield  {author} {\bibinfo {author} {\bibfnamefont {J.}~\bibnamefont
  {Wang}}, \bibinfo {author} {\bibfnamefont {J.}~\bibnamefont {Jing}}, \ and\
  \bibinfo {author} {\bibfnamefont {H.}~\bibnamefont {Fan}},\ }\href {\doibase
  10.1103/PhysRevD.90.025032} {\bibfield  {journal} {\bibinfo  {journal} {Phys.
  Rev. D}\ }\textbf {\bibinfo {volume} {90}},\ \bibinfo {pages} {025032}
  (\bibinfo {year} {2014})}\BibitemShut {NoStop}%
\bibitem [{\citenamefont {Wu}\ and\ \citenamefont {Zeng}(2022)}]{Wu2022}%
  \BibitemOpen
  \bibfield  {author} {\bibinfo {author} {\bibfnamefont {S.~M.}\ \bibnamefont
  {Wu}}\ and\ \bibinfo {author} {\bibfnamefont {H.~S.}\ \bibnamefont {Zeng}},\
  }\href {\doibase 10.1140/epjc/s10052-022-10679-1} {\bibfield  {journal}
  {\bibinfo  {journal} {Eur. Phys. J. C}\ }\textbf {\bibinfo {volume} {82}},\
  \bibinfo {pages} {716} (\bibinfo {year} {2022})}\BibitemShut {NoStop}%
\bibitem [{\citenamefont {Haddadi}\ \emph {et~al.}(2024)\citenamefont
  {Haddadi}, \citenamefont {Yurischev}, \citenamefont {Abd-Rabbou},
  \citenamefont {Azizi}, \citenamefont {Pourkarimi},\ and\ \citenamefont
  {Ghominejad}}]{Haddadi2024}%
  \BibitemOpen
  \bibfield  {author} {\bibinfo {author} {\bibfnamefont {S.}~\bibnamefont
  {Haddadi}}, \bibinfo {author} {\bibfnamefont {M.~A.}\ \bibnamefont
  {Yurischev}}, \bibinfo {author} {\bibfnamefont {M.~Y.}\ \bibnamefont
  {Abd-Rabbou}}, \bibinfo {author} {\bibfnamefont {M.}~\bibnamefont {Azizi}},
  \bibinfo {author} {\bibfnamefont {M.~R.}\ \bibnamefont {Pourkarimi}}, \ and\
  \bibinfo {author} {\bibfnamefont {M.}~\bibnamefont {Ghominejad}},\ }\href
  {\doibase 10.1140/epjc/s10052-024-12393-6} {\bibfield  {journal} {\bibinfo
  {journal} {Eur. Phys. J. C}\ }\textbf {\bibinfo {volume} {84}},\ \bibinfo
  {pages} {42} (\bibinfo {year} {2024})}\BibitemShut {NoStop}%
\bibitem [{\citenamefont {Bell}(1964)}]{Bell1964}%
  \BibitemOpen
  \bibfield  {author} {\bibinfo {author} {\bibfnamefont {J.~S.}\ \bibnamefont
  {Bell}},\ }\href@noop {} {\bibfield  {journal} {\bibinfo  {journal}
  {Physics}\ }\textbf {\bibinfo {volume} {1}},\ \bibinfo {pages} {195}
  (\bibinfo {year} {1964})}\BibitemShut {NoStop}%
\bibitem [{\citenamefont {{De Fabritiis}}\ \emph {et~al.}(2023)\citenamefont
  {{De Fabritiis}}, \citenamefont {Roditi},\ and\ \citenamefont
  {Sorella}}]{DeFabritiis2023}%
  \BibitemOpen
  \bibfield  {author} {\bibinfo {author} {\bibfnamefont {P.}~\bibnamefont {{De
  Fabritiis}}}, \bibinfo {author} {\bibfnamefont {I.}~\bibnamefont {Roditi}}, \
  and\ \bibinfo {author} {\bibfnamefont {S.~P.}\ \bibnamefont {Sorella}},\
  }\href {\doibase 10.1016/j.physletb.2023.138198} {\bibfield  {journal}
  {\bibinfo  {journal} {Phys. Lett. B}\ }\textbf {\bibinfo {volume} {846}},\
  \bibinfo {pages} {138198} (\bibinfo {year} {2023})}\BibitemShut {NoStop}%
\bibitem [{\citenamefont {{De Fabritiis}}\ \emph {et~al.}(2024)\citenamefont
  {{De Fabritiis}}, \citenamefont {Guedes}, \citenamefont {Guimaraes},
  \citenamefont {Roditi},\ and\ \citenamefont {Sorella}}]{DeFabritiis2024}%
  \BibitemOpen
  \bibfield  {author} {\bibinfo {author} {\bibfnamefont {P.}~\bibnamefont {{De
  Fabritiis}}}, \bibinfo {author} {\bibfnamefont {F.~M.}\ \bibnamefont
  {Guedes}}, \bibinfo {author} {\bibfnamefont {M.~S.}\ \bibnamefont
  {Guimaraes}}, \bibinfo {author} {\bibfnamefont {I.}~\bibnamefont {Roditi}}, \
  and\ \bibinfo {author} {\bibfnamefont {S.~P.}\ \bibnamefont {Sorella}},\
  }\href {\doibase 10.1103/PhysRevD.109.045020} {\bibfield  {journal} {\bibinfo
   {journal} {Phys. Rev. D}\ }\textbf {\bibinfo {volume} {109}},\ \bibinfo
  {pages} {045020} (\bibinfo {year} {2024})}\BibitemShut {NoStop}%
\bibitem [{\citenamefont {Guimaraes}\ \emph {et~al.}(2024)\citenamefont
  {Guimaraes}, \citenamefont {Roditi}, \citenamefont {Guedes},\ and\
  \citenamefont {Sorella}}]{Guimaraes2024}%
  \BibitemOpen
  \bibfield  {author} {\bibinfo {author} {\bibfnamefont {M.~S.}\ \bibnamefont
  {Guimaraes}}, \bibinfo {author} {\bibfnamefont {I.}~\bibnamefont {Roditi}},
  \bibinfo {author} {\bibfnamefont {F.~M.}\ \bibnamefont {Guedes}}, \ and\
  \bibinfo {author} {\bibfnamefont {S.~P.}\ \bibnamefont {Sorella}},\
  }\href@noop {} {\bibfield  {journal} {\bibinfo  {journal} {J. High Energy
  Phys.}\ }\textbf {\bibinfo {volume} {2024}},\ \bibinfo {pages} {31} (\bibinfo
  {year} {2024})}\BibitemShut {NoStop}%
\bibitem [{\citenamefont {Barnett}\ and\ \citenamefont
  {Radmore}(2002)}]{Barnett2002}%
  \BibitemOpen
  \bibfield  {author} {\bibinfo {author} {\bibfnamefont {S.}~\bibnamefont
  {Barnett}}\ and\ \bibinfo {author} {\bibfnamefont {P.~M.}\ \bibnamefont
  {Radmore}},\ }\href@noop {} {\emph {\bibinfo {title} {Methods in theoretical
  quantum optics}}},\ Vol.~\bibinfo {volume} {15}\ (\bibinfo  {publisher}
  {Oxford University Press},\ \bibinfo {year} {2002})\BibitemShut {NoStop}%
\bibitem [{\citenamefont {Clauser}\ \emph {et~al.}(1969)\citenamefont
  {Clauser}, \citenamefont {Horne}, \citenamefont {Shimony},\ and\
  \citenamefont {Holt}}]{Clauser1969}%
  \BibitemOpen
  \bibfield  {author} {\bibinfo {author} {\bibfnamefont {J.~F.}\ \bibnamefont
  {Clauser}}, \bibinfo {author} {\bibfnamefont {M.~A.}\ \bibnamefont {Horne}},
  \bibinfo {author} {\bibfnamefont {A.}~\bibnamefont {Shimony}}, \ and\
  \bibinfo {author} {\bibfnamefont {R.~A.}\ \bibnamefont {Holt}},\ }\href
  {\doibase 10.1103/PhysRevLett.23.880} {\bibfield  {journal} {\bibinfo
  {journal} {Phys. Rev. Lett.}\ }\textbf {\bibinfo {volume} {23}},\ \bibinfo
  {pages} {880} (\bibinfo {year} {1969})}\BibitemShut {NoStop}%
\bibitem [{\citenamefont {Brunner}\ \emph {et~al.}(2014)\citenamefont
  {Brunner}, \citenamefont {Cavalcanti}, \citenamefont {Pironio}, \citenamefont
  {Scarani},\ and\ \citenamefont {Wehner}}]{Brunner2014}%
  \BibitemOpen
  \bibfield  {author} {\bibinfo {author} {\bibfnamefont {N.}~\bibnamefont
  {Brunner}}, \bibinfo {author} {\bibfnamefont {D.}~\bibnamefont {Cavalcanti}},
  \bibinfo {author} {\bibfnamefont {S.}~\bibnamefont {Pironio}}, \bibinfo
  {author} {\bibfnamefont {V.}~\bibnamefont {Scarani}}, \ and\ \bibinfo
  {author} {\bibfnamefont {S.}~\bibnamefont {Wehner}},\ }\href {\doibase
  10.1103/RevModPhys.86.419} {\bibfield  {journal} {\bibinfo  {journal} {Rev.
  Mod. Phys.}\ }\textbf {\bibinfo {volume} {86}},\ \bibinfo {pages} {419}
  (\bibinfo {year} {2014})}\BibitemShut {NoStop}%
\bibitem [{\citenamefont {Ye}\ \emph {et~al.}(2016)\citenamefont {Ye},
  \citenamefont {Wang},\ and\ \citenamefont {Fei}}]{Ye2016}%
  \BibitemOpen
  \bibfield  {author} {\bibinfo {author} {\bibfnamefont {B.-L.}\ \bibnamefont
  {Ye}}, \bibinfo {author} {\bibfnamefont {Y.-K.}\ \bibnamefont {Wang}}, \ and\
  \bibinfo {author} {\bibfnamefont {S.-M.}\ \bibnamefont {Fei}},\ }\href
  {\doibase 10.1007/s10773-015-2862-1} {\bibfield  {journal} {\bibinfo
  {journal} {Int. J. Theor. Phys.}\ }\textbf {\bibinfo {volume} {55}},\
  \bibinfo {pages} {2237} (\bibinfo {year} {2016})}\BibitemShut {NoStop}%
\bibitem [{\citenamefont {Huang}(2014)}]{Huang2014}%
  \BibitemOpen
  \bibfield  {author} {\bibinfo {author} {\bibfnamefont {Y.}~\bibnamefont
  {Huang}},\ }\href {\doibase 10.1088/1367-2630/16/3/033027} {\bibfield
  {journal} {\bibinfo  {journal} {New J. Phys.}\ }\textbf {\bibinfo {volume}
  {16}},\ \bibinfo {pages} {033027} (\bibinfo {year} {2014})}\BibitemShut
  {NoStop}%
\bibitem [{\citenamefont {Ye}\ and\ \citenamefont {Fei}(2016)}]{Ye2016a}%
  \BibitemOpen
  \bibfield  {author} {\bibinfo {author} {\bibfnamefont {B.-L.}\ \bibnamefont
  {Ye}}\ and\ \bibinfo {author} {\bibfnamefont {S.-M.}\ \bibnamefont {Fei}},\
  }\href@noop {} {\bibfield  {journal} {\bibinfo  {journal} {Quantum Inf.
  Process.}\ }\textbf {\bibinfo {volume} {15}},\ \bibinfo {pages} {279}
  (\bibinfo {year} {2016})}\BibitemShut {NoStop}%
\end{thebibliography}
\end{document}